\documentstyle[12pt,aasms4]{article} 
\tighten
\def\lsim{\lower.5ex\hbox{$\; \buildrel < \over \sim \;$}}
\def\gsim{\lower.5ex\hbox{$\; \buildrel > \over \sim \;$}} 
\def\lax    {\ifmmode{_<\atop^{\sim}}\else{${_<\atop^{\sim}}$}\fi}
\def\gax    {\ifmmode{_>\atop^{\sim}}\else{${_>\atop^{\sim}}$}\fi}
\def\etal{{\it et al.\/} }
\def\gtorder{\mathrel{\raise.3ex\hbox{$>$}\mkern-14mu
             \lower0.6ex\hbox{$\sim$}}}
\def\ltorder{\mathrel{\raise.3ex\hbox{$<$}\mkern-14mu
             \lower0.6ex\hbox{$\sim$}}}

\def\pmb#1{\setbox0=\hbox{#1}%
  \kern-0.015em\copy0\kern-\wd0
  \kern0.03em\copy0\kern-\wd0
  \kern-0.015em\raise0.0433em\box0 }

\begin{document}

\title{Mechanisms for High-frequency QPOs in Neutron Star 
and Black Hole Binaries}

\author{Lev Titarchuk} 
\affil{Laboratory for High Energy Astrophysics, Goddard Space 
Flight Center, Greenbelt MD 20771, and George Mason University/CSI;
titarchuk@lheavx.gsfc.nasa.gov}

\author{Iosif Lapidus} 

\affil{Institute of Astronomy, University of Cambridge, 
Madingley Road, Cambridge CB3 0HA, UK; and Astronomy Centre,
University of Sussex, Falmer, Brighton BN1 9QJ, UK;
lapidus@star.maps.susx.ac.uk}

\author{Alexander Muslimov \footnote{NRC Resident Research Associate}} 
\affil{Laboratory for High Energy Astrophysics, Goddard Space 
Flight Center, Greenbelt MD 20771;
 muslimov@lhea1.gsfc.nasa.gov}

\vskip 0.5 truecm

\font\rom=cmr10
\centerline{\rom Accepted, Astrophys. J. }

\begin{abstract}

We explain the millisecond variability detected by Rossi X-ray
Timing Explorer (RXTE) in the X-ray emission from a number of low mass
X-ray binary systems (Sco X-1, 4U1728-34, 4U1608-522, 4U1636-536,
4U0614+091, 4U1735-44, 4U1820-30, GX5-1 and {\it etc}) 
in terms of dynamics of the centrifugal barrier, a hot boundary region 
surrounding
a neutron star (NS). We demonstrate that this region may experience
the relaxation oscillations, and that the displacements of a gas element
both in radial and vertical directions occur at the same main
frequency, of order of the local Keplerian frequency. We show  
the importance of the effect of a splitting of the main frequency 
produced by the Coriolis force in a rotating disk for the
interpretation of a spacing between the QPO peaks. We estimate a 
magnitude of the splitting effect and present a simple formula for 
the whole spectrum of the split frequencies. It is interesting that the 
first three lowest-order overtones (corresponding to the azimuthal 
numbers $\rm m = 0,~-1,~and ~-2$) fall in the range of 200-1200 Hz 
and match the kHz-QPO frequencies observed by RXTE. Similar phenomena 
should also occur in Black Hole (BH) systems, but, since the QPO 
frequency is inversely proportional to the mass of a compact object, 
the frequency of the centrifugal-barrier oscillations in the BH systems
should be a factor of 5-10 lower than that for the NS systems. The 
X-ray spectrum formed in this region is a result of upscattering of 
a soft radiation (from a disk and a NS surface) off relatively hot 
electrons in the boundary layer. The typical size of the emission
region should be 1-3 km, which is consistent with the time-lag 
measurements. We also briefly discuss some alternative QPO models, 
including a possibility of acoustic oscillations in the boundary
layer, the proper stellar rotation, and $g$-mode disk oscillations.

\end{abstract}

\keywords{accretion, accretion disks --- black hole physics 
--- radiation mechanisms: thermal ---stars:neutron --- X-rays: general}

\section{Introduction} 

The process of accretion onto a compact object in the neutron star (NS) 
and black hole (BH) binaries has many remarkable similarities. Some  
part of accreting matter with large angular momentum forms a disk, 
while another part (participating in a sub-Keplerian rotation) undergoes
practically a free fall accretion until the centrifugal barrier (CB) 
becomes sufficient to halt the flow (see e.g. Chakrabarti \& Titarchuk 
1995; hereafter CT95). Thus, one may expect that two distinct zones, 
a disk and a barrier, can be formed in the vicinity of a compact
object which are likely to be responsible for the generation of a 
resulting spectrum. In this paper we consider a possibility of the 
dynamical adjustment of a Keplerian disk to the surface of a NS. 

The  disk structure begins deflecting from a Keplerian one at a 
certain point to adjust itself to the boundary conditions 
at a NS surface (or at the last stable orbit, at $\rm R_0=3R_s$, 
in the case of a BH). The transition from a Keplerian to a 
sub-Keplerian flow may proceed smoothly, though it is very likely  
that a perfect adjustment never occurs. In general, the transition 
should take place through the setting up of the CB (where a centrifugal 
force slightly exceeds the gravitational force) within the adjustment 
radius. Here (see \S 2.1 for details) we discuss the 
formation of kinks and shocks in the supersonic regime of accretion 
flow as a possible physical reason for a super-Keplerian rotation. 
We suggest that in a region with a super-Keplerian rotation a matter may 
experience the relaxation oscillations in the vertical and radial
directions. These oscillations are expected to be in a resonance with 
the local angular velocity in a disk, and the variation in an emitting 
area caused by the oscillations around a transition point produces 
the QPOs in the X-ray flux.

The principal observational consequence of our scenario is a correlation
between the X-ray flux and QPO frequency. The relatively
soft disk photons in the CB region are scattered off the hot electrons
thus forming the Comptonized X-ray spectrum (Sunyaev \& Titarchuk 1980; 
hereafter ST80). The electron temperature is regulated by the supply
of soft photons from a disk, which depends on the ratio of 
the energy release (accretion rate) in the disk and the energy release 
in the CB region (see Eq. [24] for more details). For example, the 
electron temperature is higher for lower 
accretion rates (see e.g. CT95, Figures 3, 4), while for a high 
accretion rate (of order of the Eddington one) the CB region 
cools down very efficiently due to Comptonization.

In this paper we discuss the CB model as a possible mechanism for 
the $\sim 1$ kHz QPOs discovered by Rossi X-ray Timing Explorer (RXTE)
in a number of LMXBs (see Strohmayer \etal 1996, and Van der Klis 
\etal 1996, hereafter S96, and VK96, respectively; Zhang \etal 1996). 
These observations reveal a wealth
of previously unknown high-frequency X-ray variabilities which are
believed to be due to the processes occurring on, or in the very 
vicinity of an accreting NS.

In \S 2 we schematically describe our CB model. We
demonstrate that the main model parameter $\gamma$ (which is proportional
to a ratio of accretion rate to viscosity and is in
fact Reynolds number for the accretion flow) 
determines the QPO frequency.
As the  Reynolds number ($\gamma$-parameter) increases, the CB moves toward 
a NS surface (or toward the inner edge of an accretion disk), and 
the X-ray flux at higher energies declines because of the cooling 
of the emission region (CB). In addition, as the  Reynolds 
number ($\gamma$-parameter) increases, the QPO frequency reaches its limiting 
value (\S 2.1). We estimate the characteristic frequency of the 
oscillations of the CB region in \S 2.2. In \S 2.3, we show that 
due to rotation of a disk the effect of the Coriolis force should 
result in the splitting of the main QPO frequency, and in \S 2.4, we 
present our estimate of the corresponding oscillation amplitude. 
We apply our CB model to the interpretation 
of RXTE observations of high-frequency QPOs in \S 3. In \S 4, we 
discuss the alternative possibilities including resonant acoustic
waves in the hot boundary layer around the NS, stellar rotation,  disk 
nonradial oscillations for the QPOs with a frequency of $\sim $300
Hz and g-mode disk oscillations. Finally, we summarize our conclusions in \S 5.

\section{Centrifugal barrier oscillations}

\subsection{The formulation of the problem}

Let us consider a Keplerian accretion disk around a BH or a
weakly magnetized NS. It is generally believed that there is a
transition layer in the vicinity of a compact object where an 
accreting matter adjusts itself either to the surface of a rotating
NS, or to the innermost boundary of an accretion
disk. In the transition layer the motion is not Keplerian, and is
governed by the mechanism of the angular momentum loss by an 
accreting matter. Thus, we may define the transition layer as a
region confined between the surface of a NS (either the last stable 
orbit or the corotating magnetosphere) and the first Keplerian orbit. 
In a Keplerian 
disk a matter is circularly orbiting with the angular velocity
$$
\rm \omega_K=\left({{GM}\over{R^3}}\right)^{1/2}, 
\eqno(1)
$$
where G is the gravitational constant, M is the mass of a
compact object, and R is the radius of an orbit. The radial motion 
in a disk is provided by the friction and angular momentum exchange 
between the adjacent layers that result in the loss of initial angular
momentum by an accreting matter. The radial transport of the angular 
momentum in a disk can be written therefore in terms of the torque of 
viscous forces between the adjacent layers (see e.g. Shakura 
\& Sunyaev 1973)
$$
\rm \dot M {d\over {dR}}(\omega R^2) = 
2\pi {d\over {dR}} (W_{r\varphi}R^2) . \eqno(2)
$$
Here $\rm \dot{M}$ is the accretion rate, and $\rm W_{r\varphi}$ is 
the component of a viscous stress tensor that can be expressed as
$$
\rm W_{r\varphi}=-2\eta H_dR{{d\omega}\over{dR}}, 
\eqno(3)
$$
where $H_d$ is a half-thickness of a disk, and $\eta$ is the turbulent
viscosity. The only parameter entering this equation is the ratio
$$
\rm \gamma={{\dot M}\over{4\pi\eta H_d}}
%={{3R v_r}%\over {{\it v}_t{\it l}_t}},
\eqno(4)
$$
which is nothing else but Reynolds number for the accretion flow. 
The viscosity determines the redistribution of the momentum in the
flow, with both the particles and photons (in the radiation dominated
region) participate in momentum transport between the shearing
layers. Narayan (1992) has argued  that the momentum transport
(viscous diffusion) by the particles in the flow depends on  
the velocity of the flow if the sources supplying particles 
are ``frozen'' into the flow. Specifically, Narayan (1992) has
demonstrated that if the particle sources comove with the flow at the
bulk velocity $V$, and the particles diffuse with the
velocity $c_s$, then the effective diffusion (viscosity) 
coefficient scales as $(1-V^2/c_s^2)$. 
%For a simple case of the one-dimensional particle
%transport Narayan (1992) has shown that the diffusion approximation
%leads to the unphysical result violating causality. He compared the
%exact result with the diffusion-approximation solution and found that 
%the upstream branch of the exact solution vanishes as $V\rightarrow
%c_s$, which is physically reasonable and satisfies the causality principle. 
%The similar result has been obtained for the two-dimensional case as well.
Thus, the viscosity of the supersonic flow gets
effectively suppressed provided that the collective plasma effects 
(Tsytovich 1977) do not have enough time to develop, which seems to be 
the case in the limit of $V \approx c_s$ where the distribution of particles in
the flow is essentially determined by the source (see Narayan 1992,
Equation [3.5]). 

The very important consequence of the Narayan's
result is that the regime of high $\gamma$ (Reynolds number) can
occur even for the low accretion rates, $\dot M$, in the presence
of high velocities of the flow. However, the radiative viscosity 
can be important under the radiation-pressure dominated conditions and 
in the presence of high velocities of the flow, which seems to be
the case for the transition layer.
The coefficient of radiative viscosity is  given by 
$$
\rm \eta={1\over3}{\rm m_p n_{ph}} {\it c}{\it l},
\eqno(5)
$$
where $\rm m_p$ is the proton mass, $\rm n_{ph}$ is the photon number 
density, $\it c$ is the speed of light, and $\rm {\it l}$ is the 
photon mean-free path. The photon number density is proportional 
to the accretion rate, $\dot M$. 
%However, because of the
%dependence of the photon mean-free path $ {\it l}$ on the accretion 
%rate, the $\gamma-$parameter is now a decreasing function of 
%$\dot M$. 
Thus, it is important that if the KHz QPO 
phenomenon is mostly determined by the ratio of accretion rate to 
the radiative viscosity of the flow, $\dot M/\eta$, then we would 
expect it to occur in the sources with both
high and low accretion rates.

%where $\rm {\it v}_t$ and $\rm {\it l}_t$ are the velocity 
%and length-scale of turbulent eddies. 
 
%Within the framework of Shakura-Sunyaev model 

We can calculate the
distribution of angular velocity of matter in the transition layer by
solving equation (2) with the appropriate boundary conditions.  We
adopt that at the inner boundary the angular velocity of matter either
matches the rotation velocity of a NS (or velocity of the corotating 
magnetosphere at the magnetopause),
$$
\rm \omega=\omega_0~~~~~~{\rm~at}~~~~~~ R=R_0,
\eqno(6a)
$$
or vanishes (e.g. in the case of a non-rotating BH), 
$$
\rm \omega=0~~~~~~{\rm~at}~~~~~~ R=R_0.
\eqno(6b)
$$
The outer boundary condition for the layer must require that the orbital
motion of matter matches smoothly a Keplerian motion at some radius
$\rm R_{out}$. The latter means that the angular velocity of matter 
and its radial derivative should be equal to those 
for a Keplerian disk at the same radius, i.e.
$$
\rm \omega=\omega_K~~~~~~at~~~~~~\rm R=R_{out},
\eqno(7)
$$
and 
$$
\rm {{d\omega}\over{dr}}={{d\omega_k}\over{dr}}~~~~~~
at~~~~~~\rm R=R_{out}.
\eqno(8)
$$
The boundary conditions (6)-(8) allow us to solve equation (2). Thus, 
we can unambiguously calculate a radial profile of the angular
velocity in a transition layer $\rm \omega(R)$, and derive  the value
of the outer radius, $\rm R_{out}$. It is important, that the only 
parameter that controls the adjustment of the angular velocity in 
a transition layer to the Keplerian angular velocity in a disk is 
$\gamma-$parameter.
%where $\rm v_r$ is the radial component of velocity.   

%We must note that equation (2) is the second-order differential
%equation, and the fact that $\gamma$ may actually depend on radial
%coordinate does not affect our qualitative conclusions. Thus, for the
%illustration purposes and without loss of generality we shall present 
%the solution where $\gamma $ is a constant. 
Let us focus on how to 
get a formal solution that would describe a smooth transition of the 
Keplerian flow into the sub-Keplerian rotation. The corresponding 
problem for equation (2) should be formulated as the boundary problem 
for the second-order differential equation with the {\it three}
boundary conditions: one at the inner boundary, and two at the
adjustment point, where the flow begins to deviate from the Keplerian 
motion in order to adjust itself to the sub-Keplerian rotation.  
Thus, we have {\it three} boundary conditions for
the {\it second} order differential equation which implies that 
the position of the adjustment point can be determined uniquely 
(see e.g. Korn \& Korn 1961, Ch. 9.3). Such a simple formulation is 
quite instructive: it guarantees  the uniqueness
of the solution and automatically fixes the position of the outer
boundary. This is important because the position of the outer boundary
of the transition layer is {\it a priory} not known. Due to this
very general mathematical reasoning the specific radial dependence of the
coefficients in the differential equation (2) is not of principal
importance.
                                                                         
%Since we are interested in a region close enough to the inner boundary
%(i.e. $\rm R_{out}-R_0\ll R_0$), in the first approximation 
To illustrate how one can fix the position of an adjustment radius, 
we shall present the solution of equation (2) for the case where the 
$\gamma$-parameter is a constant.  Let us now introduce the dimensionless
variables: angular velocity $\theta=\omega/\omega_0$, radius $\rm r=R/R_0$
($\rm R_0=3R_s$, $\rm R_s=2GM/c^2$ is the Schwarzschild radius), and mass
$\rm m=M/M_{\odot}$. We
 normalize the angular frequency by the value $\rm \omega_0=2\pi
\times 363~rad \cdot s^{-1}$.  The main reason for this is the remarkable
363-Hz QPO discovered during the type-I X-ray bursts from 4U1728-34 (VK96), 
and which is believed to be associated with 
the rotation frequency of a NS.
  
In terms of these variables the Keplerian angular
velocity reads
$$
\rm \theta_K={{6}\over{m r^{3/2}}}.
\eqno(9)
$$ 
The solution of equation (2) satisfying the boundary conditions 
(5)-(7) is
$$
\rm \theta(r)=D_1 r^{-\gamma} + (C_{NS}-D_1) r^{-2},
\eqno(10)
$$
where 
$$
\rm D_1={{\theta_{out}-C_{NS}r_{out}^{-2}}\over{r_{out}^{-\gamma}
-r_{out}^{-2}}}~~~~~and~~~~~\theta_{out}=\theta_K(r_{out}).
$$
The value of the dimensionless
outer radius, $r_{out}$, should be determined from equation (7), which can
be now rewritten as
$$
\rm {3\over2} \theta_{out}=D_1 \gamma r_{out}^{-\gamma}+2(C_{NS}
-D_1)r_{out}^{-2}.
\eqno(11)
$$
Thus, equations (10) and (11) represent the solution comprising both 
a BH ($\rm C_{NS}=0$) and a NS ($\rm C_{NS}=1$) cases. 
\par
\noindent
{\bf Editor, Please put Fig.1 here}
\par
\noindent
In Figure 1 
we plot a family of solutions of equation (2) for a NS case. The solid
line shows a Keplerian solution described by formula (9), the dashed 
line shows the solution with a sub-Keplerian rotation in a transition 
region, and the dash-dotted line shows the solution with a 
super-Keplerian rotation in the upper half of a transition region. 
The formal explanation for these types of solutions is the fact that 
the occurrence of kinks in an accretion flow enhances (with respect to
the Keplerian value) the absolute value of a radial derivative of the 
angular velocity at the outer boundary. {\it The perfect adjustment of 
the Keplerian rotation to the sub-Keplerian inner boundary condition
(see equation [10]) is possible only in a rather unrealistic case where at
the adjustment point the angular velocity and its radial derivatives 
coincide with the corresponding Keplerian values. In this case one can
uniquely determine the position of an adjustment point. However, if 
 this continuous outer boundary condition (see [6] and [7]) is 
not fulfilled, then the solution of equation (2) subject to the 
inner sub-Keplerian boundary condition should necessarily have a
regime corresponding to the super-Keplerian rotation}.

It is important, that the kinks generally imply the presence of 
the discontinuities in a radial 
derivative of an angular velocity of the flow. Thus, when this 
derivative exceeds the corresponding Keplerian value, the adjustment 
of a flow to the inner boundary condition most likely occurs through 
the formation of a region with a locally super-Keplerian rotation. 
The physical reason for a super-Keplerian rotation in a transition 
region is a partial outward transport of the angular momentum (e.g. 
facilitated by the formation of kinks in a flow) due to viscous 
stresses.  As a result of this redistribution of the angular 
momentum, the matter, before entering a regime of an essentially 
sub-Keplerian rotation (in the innermost part of a transition region),
gets locally involved in a super-Keplerian rotation. 

Let us discuss the reasoning for the occurrence of a super-Keplerian 
rotation in more detail. It is well-known from the standard accretion 
theory (see e.g. Shakura \& Sunyaev 1973, Lipunov 1992, 
Popham \& Narayan 1992) that, depending on the value 
of the viscosity (see Popham \& Narayan 1992 for the detailed
discussion of the viscosity effects), 
the flow may be either subsonic 
or supersonic. When the flow reaches a supersonic velocity the 
density of matter decreases (see Landau \& Lifshitz 1988, 
equation [83,5]), and a cooling regime changes from the free-free 
cooling onto the Compton cooling (see \S 2.3, and also Zel'dovich \& 
Shakura 1969). As a result, the temperature of matter increases, 
the effects of radiative conduction overtake the effects of
hydrodynamic viscosity, and a strong isothermal shock develops
(Landau \& Lifshitz 1988, \S 95). Note also, that the change in the 
curvature of an accretion disk caused by an asymmetric illumination of
a disk by a NS (Maloney, Begelman \& Pringle 1996) favors the
formation of kinks. The kinks are characterized by the
discontinuities in the coordinate derivatives of the velocity
components. In their discussion of kinks in the hydrodynamic flows 
Landau \& Lifshitz (1988, \S 96) emphasize a difference between the 
formation of shocks and kinks: the shocks are necessarily formed due 
to the specific continuous boundary conditions, while the kinks are 
always a result of singularities in the boundary or initial
conditions. Thus, in the presence of kinks the adjustment of a  
Keplerian flow to the inner boundary condition can occur through 
a super-Keplerian motion.

The Keplerian rotation can be adjusted to 
the condition at the inner boundary (see Figure 1) only if the
absolute value of a radial derivative of angular velocity at the 
radius of kink formation exceeds the corresponding Keplerian value. When a 
super-Keplerian motion occurs a matter piles up in the 
vertical direction thus disturbing the hydrostatic equilibrium. The 
restoring force due to the vertical component of the gravitational 
force prevents matter from further accumulation in a vertical
direction and supports relaxation oscillations. The radiation drag
force, which is proportional to the vertical velocity component, 
determines the characteristic decay time of the vertical oscillations.

In our calculations illustrated in Figure 1 the dimensionless 
thickness of a region with a super-Keplerian rotation is about 
0.2 ($=2.4$ km, for a canonical NS of 1.4 M$_{\odot}$ mass and 
10 km radius). The interpretation of the $\sim 27~\mu$s time lags 
between the softer and harder components of a spectrum observed 
in kHz QPO from 4U 1636-53 (Vaughan \etal 1997) in terms of 
the ST80 Comptonization model gives a similar estimate for a 
size of the hard emission region. The development of a zone with 
a super-Keplerian rotation may significantly affect the dynamics 
of a transition region: the CB may be set up on the way of a matter 
accreting onto a compact object. Another effect we shall address in
this paper is a possibility of global oscillations of a ring-like
configuration (see Figure 2 for an artistic concept of our model) 
formed by a matter accumulated near the CB. We will discuss these 
oscillations in Section 2.2 in a very schematic and semi-quantitative 
way. The detailed modeling of the transition region and analysis of 
its oscillation properties in the presence of the CB will be presented 
in a separate publication. We will also consider the effects of 
rotation on the oscillations in the case of an axisymmetric 
equilibrium disk structure.

{\bf Editor, Please put Fig.2 here}

\subsection{QPO frequency of the centrifugal barrier region}

Figure 1 shows that in the transition region there is 
a layer where the centrifugal force exceeds the gravitational force. 
This means that in this layer  matter tends to pile up before it 
adjusts itself to a new hydrostatically equilibrium structure in the 
vertical direction. It is very likely that the relaxation to a new 
equilibrium structure will be accompanied by the nonradial 
oscillations of a layer. The detailed spectrum of oscillation modes 
depends on the physical parameters of the unperturbed configuration 
and is a subject of a special analysis (see e.g. Appendix B for a 
general introduction). In this paper, we would like to present an 
order-of-magnitude estimate based on a rather simplified model. Let 
us consider the small displacement, $\rm h(R,t)$, of the gas element in a 
vertical direction at the radius R. We assume that
in  hydrostatic equilibrium, the vertical gradient of the pressure P
($\rm = P_g+P_{rad}$, where $\rm P_g=2\rho kT/m_p$ is the gas
pressure, and $\rm P_{rad}=\sigma_T\rho H_d L/m_p$ is the radiation
pressure) is balanced by the vertical component of the gravity force 
$\rm F_g\approx \rho (GM/R^2)H_d(H_d/R)$.
Thus, the equilibrium condition for a small vertical displacement
reads (neglecting general-relativistic corrections) 
$$
\rm \rho GMH_d^2/R^3 = \sigma_T\rho H_d L/m_p+2\rho kT/m_p,
\eqno(12)
$$
where L(r) is the local radiation flux per cm$^2$. Note that in this
equation we have neglected the vertical component of a centrifugal 
force which is very small compared to the vertical component of a 
gravitational force.

A small vertical displacement of a gas layer from the equilibrium will
result in the restoring force $\rm f_{gr} \approx -
m_pGMh/R^3$ and the radiation drag force $\rm |f_{r}|
\ltorder (4/3) \ell [GMH_d)/R^3](\dot h/c)$. Here $\rm
\ell=L/L_{Ed}$ is the luminosity in Eddington units, and the
inequality holds for a nonblackbody radiation field (the equality 
holds for the blackbody radiation field only). Thus, neglecting
the density and pressure perturbations, we can write the equation
describing the vertical oscillations of a layer (the equation of
motion)
$$
\rm m_p\ddot h= f_{gr} - |f_r| ,
\eqno(13)
$$
where dot means a time derivative. Now, by making use of 
equations (12) and (13), we can rewrite the oscillation equation in the form
$$
\rm \ddot h + \ell[GMH_d/cR^3]\dot h+[GM/R^3]h=0.
\eqno(15)
$$
For a harmonic perturbing force ($\rm \propto {\it e}^{i\Omega t}$) 
the power spectrum of the oscillations reads 
$$
\rm <P(\Omega)> \,\propto{{\gamma_{d}}\over{(\Omega^2-\Omega_0^2)^2+
\gamma_{d}^2\Omega^2}},
\eqno(16)
$$
where $\rm \Omega_0^2=GM/R^3$ is the eigen-frequency 
of the oscillator, and
$\rm \gamma_d= \ell \Omega_0^2 H_d/c$ is the damping rate of
oscillations. We can now estimate the oscillator's Q-value:
$$
\rm Q=\Omega_0/\gamma_d =c/(H_d\ell\Omega_0)\gtorder 300 .
\eqno(17)
$$
The emission temperature is determined by the accretion rate 
(see \S 2.3), and therefore the Q-value which scales as 
$$
\rm Q=c/(H_d\ell\Omega_0)= \left({{m_pc^2}\over{kT}}\right)^
{1/2}{{1}\over{\ell}},
\eqno(18)
$$
should also depend (through the temperature) on the accretion
rate. Namely, {\bf the Q-value should increase with 
increasing a centroid frequency of the QPO, since the temperature of 
an oscillating layer decreases as the accretion rate increases.}

{\bf Editor, Please put Fig.3 here}

In Figure 3 we present the QPO frequency  as a
function of the $\gamma$-parameter. This Figure implies that the QPO
frequency should scale with the $\gamma-$parameter 
%(i.e. on the 
%accretion rate, or on the disk blackbody radiation flux) 
approximately as $\gamma ^{-0.4}$. Perhaps, this dependence manifests 
itself in the recent observational data on 4U 0614+091 (Ford \etal
1997), indicating that the main QPO frequency depends on the blackbody flux. 
.

\subsection{Temperature of a CB emission region}

The total count rate from the source increases with an increase 
in an accretion rate. This is accompanied by an increase in
the observed QPO frequency 
(if the viscosity is not affected by the radiation, see \S 2.1).
 At the same time, the X-ray
spectrum becomes softer, because of an increase in a supply of 
soft photons from the disk illuminating the emission region of the CB.

The CB region can be treated as a potential wall at which the
accreting matter releases its gravitational energy. {\it This occurs
in an optically thin region where the column density is of order of a
few grams}, or where the Thomson optical thickness $\tau_0 \sim $ a
few. The amount of energy released per second is a fraction of the Eddington
luminosity since the CB region is located in the very vicinity of
a central object ($\sim $ 3-6 $\rm R_S$). The heating of a gas due to
the gravitational energy release should be balanced by the photon
emission. For the high gas temperatures, Comptonization is the
main cooling channel, and the heating of electrons is due to their  
Coulomb collisions with protons. Under such physical conditions the
energy balance can be written as (see e.g. Zel'dovich \& Shakura 1969,
hereafter ZS69, equation [1.3])
$$
\rm F/\tau_0\sim C_0\cdot \varphi(\alpha)
\varepsilon(\tau)T_e/f(T_e) .
\eqno(19)
$$
Here $\tau$ is the current Thomson optical depth in the emission
region (e.g. in a slab), $\alpha $ is the energy spectral index for  
a power-law component of the Comptonization spectrum, 
$\varepsilon(\tau)$ is a distribution function for the radiative 
energy density, $\rm f(T_e) = 1+2.5(kT_e/m_ec^2)$, $\rm T_e$ is a 
plasma temperature in K, $\rm C_0 = 20.2$ cm s$^{-1}$ K$^{-1}$ is a 
dimensional constant, and, finally, $\rm \varphi (\alpha ) =
0.75\alpha (1+\alpha /3)$ if $\varphi (\alpha )\leq 1$, otherwise 
$\varphi (\alpha ) = 1$. The latter formula is obtained by using 
the relationship between the zero- and first-order moments (with 
respect to energy) of the Comptonized radiation field 
(Sunyaev \& Titarchuk 1985, \S 7.3, equation [30]). The distribution 
of the radiative energy density in the emission region 
$\varepsilon (\tau )$ can be obtained from the solution of 
the diffusion equation (cf. ZS69, equation [1.4]), 
$$
\rm {1\over3}{{d^2 \varepsilon}\over{d\tau^2}}=-{(F/c)\over{\tau_0}} ,
\eqno(20)
$$
subject to the two appropriate boundary conditions. 

\noindent
The first boundary condition must imply that there is no
scattered radiation from the outer side of the emission region, i.e.
$$
\rm {{d\varepsilon}\over{d\tau}}-{3\over2}\varepsilon=0~~~for~\tau=0. 
\eqno(21)
$$
In our case this condition holds at the inner surface of a slab facing
a central object. We  note that, in reality, one can expect some
additional soft flux from a NS resulting in an additional illumination 
of the inner surface of a slab. In the following analysis we shall 
neglect this effect. 

\noindent
The second boundary condition requires that at the outer surface of a
slab the incoming flux should be equal to the external flux H
(not to be confused with unrelated quantity $H_d$ which is a half-thickness
of the disk) , i.e. 
$$
\rm {1\over3}{{d\varepsilon}\over{d\tau}}={H\over c} ~~~at~\tau=\tau_0.
\eqno(22)
$$
The solution of equations (20)-(22) provides us with the distribution 
function for the energy density 
$$
\rm \varepsilon(\tau)= {{F+H}\over{c}}
\{2+3\tau_0[\tau/\tau_0-0.5(\tau/\tau_0)^2 F/(F+H)]\}.
\eqno(23)
$$
Thus, from equations (19) and (23) we get
$$
\rm \varphi(\alpha)T_e\tau_0/f(T_e)\ltorder 
0.75\cdot 10^9{{F}\over{F+H}}~{\rm K}. 
\eqno(24)
$$   
When $\rm H<F$ the spectral index $\alpha $ varies very little since
$\alpha $ is a function of $\rm T_e\tau_0/f(T_e)$ (see e.g. 
Titarchuk \& Lyubarskij 1995 for the case $\tau_0\lsim 2$). Thus, 
as long as the external
flux (due to the photons from the disk) is much smaller than the
internal energy release (per cm$^2$ per second) in the CB region, the
spectral index is insensitive to the accretion rate in the disk.  The
values of parameters consistent with equation (24) are:
$\rm \tau_0\lsim 5$, $\rm T_e\lsim 2\times 10^8$ K, and
$\alpha \lsim 1$, which are characteristic of a hard state for the 
galactic BH and NS systems.
 
When the external flux H becomes comparable to the internal energy
release, F, the cooling becomes more efficient not only due to
Comptonization, but also due to the free-free cooling, and therefore
the electron temperature unavoidably decreases, $\rm T_e\ll 10^9\cdot
F/(F+H)$ K (see CT95 for the numerical calculations of spectral
indices and temperature).  

{\bf Editor, Please put Fig.4 here}

In Figure 4 we present the plot of a
temperature in the CB emission region as a function of the H/F
ratio. The latter is proportional to the mass accretion rate in a
disk. We can calculate the temperature using expression (24), then 
we get the following formula for the spectral index 
(Sunyaev \& Titarchuk 1980)
$$
\rm \alpha=(2.25+\gamma_c)^{1/2}-1.5,
\eqno(25)
$$ 
where $\rm \gamma_c=\pi^2m_ec^2/[3 kT_e(\tau_0+2/3)^2f(T_e)]$
(not to be confused with  unrelated quantity $\gamma$ which 
is a Reynolds number of the accreting flow in  the disk).
\par
In the end of this paragraph we would like to present a rather elegant 
and instructive justification for the existence of a so-called ``hot''
solution for the case of a large optical depth ($\tau _0\gg 1$)  
of the energy release region. We must recall that a ``hot'' solution 
was discussed by Turolla et al. (1994) and Zane, Turolla \& 
Treves (1997, hereafter ZTT97) in the context of a formation and 
structure of a static atmosphere around a non-magnetized steadily 
and spherically accreting NS. The statement regarding the existence of 
a ``hot'' solution (Turrola et al. 1994) was quite surprising, and 
this solution has been even interpreted as an artifact resulted from 
a perfect-reflection inner boundary condition adopted by the authors.
Here we shall demonstrate that this solution follows directly
from our equations (19), (23), and (25) and assumption that 
$\rm H=0$ at $\tau =\tau _0\gg 1$ (the absence of an incoming flux 
from a boundary).

After substitution of (23) into equation (19) and in the limit of 
$\tau_0\gg 1$ we get
$$
\rm f^2(T_e)\approx 2.5 .
\eqno(26)
$$
This means that $\rm kT_e=120$ keV which perfectly matches 
the temperatures presented in ZTT97 (see Figure 2, there). In our derivation 
of expression (26) we have used the relation 
$\rm \alpha=\gamma_c/3$ (see equation [25] for $\alpha\ll 1$), 
and we have calculated 
the energy density, $\varepsilon (\tau)$ (see equation [23]), 
for $\tau /\tau_0 \geq 1/2$. It is very interesting that in the limit of  
$\tau_0\gg 1$ the derivation of equation (33) is practically 
independent of an optical depth of the energy release region 
$\tau_0$. Thus, we come up with exactly the same conclusion as 
in ZTT97: for a large optical depth of the energy release 
region $\tau_0$ the radiation and plasma are almost in an equilibrium,  
so that the plasma remains very hot (''hot'' solution) because of
the inefficient Compton cooling. 

\subsection{Effect of rotational splitting}

In this Section we would like to draw a reader's attention to 
the effect of rotational splitting of oscillation frequencies 
(Titarchuk \& Muslimov 1997) that was ignored in all previous 
theoretical studies of QPOs.  

It is well-known (see e.g. Unno \etal 1979 and references
therein) that in a rotating star the nonradial oscillations are
split: $\rm \sigma _{klm} = \sigma _{kl} + m \Omega C_{kl}$, where
$\rm \sigma _{kl}$ is the oscillation frequency in the non-rotating
star, $\rm \sigma _{klm}$ is the frequency in the rotating star (in
the co-rotating frame), m is the azimuthal number, $\Omega $ is the
stellar rotation frequency, and $\rm C_{kl}$ is an integral that
depends on the stellar structure and on the eigenfunction. Thus, in a
rotating star, a nonradial oscillation drifts at a rate $\rm m \Omega
C_{kl}$ relative to a fixed longitude of a star. The effect of
rotational splitting is due to the Coriolis force and is similar to
the Zeeman effect in a magnetic field.

The oscillations of a rotating disk must split as well, and the 
rotational splitting of the disk oscillations is particularly 
important for those modes whose frequencies are comparable to 
the frequency of a disk rotation relative to a distant observer. 
Our estimate (warranted when the relative correction to the 
eigenfrequency is small, $\rm \sigma ^{(1)} / 
\sigma _0 \ll 1$) of the rotational splitting for a disk (see 
Appendix A, equation [A11]) results in the following formula (see also
Titarchuk \& Muslimov 1997)
$$
\rm \Omega_{k,m}=\Omega_0 +m\left[1-{{2}
\over{1+s\pi^2k^2+m^2}}\right]\Omega,
\eqno(27)
$$
where $\rm \Omega_{k,m}$ is the oscillation frequency measured by a
distant observer, $\Omega_0$ is the eigenfrequency, $\Omega $ is the
local angular frequency of rotation of the oscillating region of a 
disk (as measured by a distant observer), $\rm s \lsim 1$ is some function
depending on a disk vertical structure (see Appendix A), and m and
k are the azimuthal and vertical mode numbers, respectively.  For the
lowest-order modes with $\rm m = 0,~-1$, and $- 2$, the oscillation
frequencies are
$$
\rm \Omega_{k,0}=\Omega_0,
\eqno(28)
$$
$$
\rm \Omega_{1,-1} =-(\Omega-\Omega_0) +{{2}\over{1+s\pi^2+1}}\Omega,
\eqno(29)
$$ 
and 
$$
\rm |\Omega_{1,-2}|=\Omega_0+2(\Omega-\Omega_0)-
{{4}\over{1+s\pi^2+4}}\Omega,
\eqno(30)
$$
respectively.

\subsection{Oscillation amplitudes}

The observed amplitudes of the flux oscillations can be explained in terms
of a variable area of the surface of a region emitting the hard
radiation. The latter is produced by the upscattering of soft
radiation from a disk. The amount of  soft radiation from a disk
intercepted by the emission region, $\rm L_s$, is proportional to its
surface area, $\rm 2\pi H_dR_{cb}$. The luminosity of  hard radiation,
$\rm L_{h}= A(\tau_0, T_e)L_s\propto A(\tau_0,T_e)2\pi H_dR_{cb}$, where
$\rm A(\tau_0,T_e)$, the enhancement factor due to Comptonization,
 is determined by the optical depth, $\tau_0$, of the CB-emission
region and by the electron temperature, $\rm T_e$ (see e.g. Titarchuk
1994).  Note that the CB oscillations do not affect the enhancement
factor, since both the optical depth (or column density) and the
temperature remain constant during the oscillations, so that their
product $\rm \tau_0 T_e\approx const$ (see equation [24]). We also
point out that the radial and vertical displacements of a ring-like CB
configuration should occur at the same frequency. Assuming that the
volume of the CB configuration, $\rm V_{cb}=2\pi H_dR\Delta R $,
does not change during the oscillations, it is straightforward to obtain 
the following relation between the vertical and radial displacements, 
$\rm \delta H_d$ and $\rm \delta(\Delta R)$, respectively,
$$
\rm \delta( R)=\delta(\Delta R)\approx-{{\delta H_d}\over{H_d}}\Delta R.
\eqno(31)
$$

To obtain an order-of-magnitude estimate of a relative amplitude of
the luminosity variation due to the oscillations at the main 
frequency $\Omega _0$ we can use the formula for a differential of 
the area of an emitting surface and equation (31),
$\rm S_{CB}=2\pi RH_d$,
$$
\rm \delta S_{CB}= 2\pi(R -\Delta R) \delta H_d\approx 2\pi R \delta H_d,
\eqno(32)
$$
Then, for a relative change in the luminosity, which is proportional
to a change in the area of an emitting surface, we get the following estimate
$$
\rm \delta L/L \approx \delta H_d/H_d .
\eqno(33)
$$
 
Thus, even small variations of a height of a cylindrical CB area 
$\rm \delta H_d \sim 0.1 H_d$ can produce oscillations in the hard 
X-ray flux at the level of $\rm \delta L \sim 0.1 L$, which is of order 
of the observed ones.  

We must note that an exact 
relation between the components of a displacement vector (see
Appendix A for the notations) $\rm \hat \xi_r$, $\hat \xi_{\varphi}$, 
and $\rm \hat \xi_{z}$ depends also on the azimuthal and vertical 
mode numbers. For example, consider an individual oscillation mode 
with displacement $\rm {\pmb{$\xi $}}^{(0)}$ in an 
incompressible fluid. Then $\rm \nabla 
\cdot {\pmb{$\xi $}}^{(0)}=0$, and we
get a linear partial differential equation that determines 
a relation between $\rm \hat \xi_r$, $\rm \hat \xi_{\varphi}$, and
$\rm \hat \xi_{z}$. Thus, in general, this relation is very 
complicated and should depend on the 
azimuthal and vertical mode numbers, m and k, respectively.

The mechanism for the QPO emission we discuss in this paper 
implies that the QPO should have larger amplitudes in the hard tail of 
the spectrum that is produced by Comptonization in the 
CB region. This is in  good agreement with the observations of 
the kHz QPO from 4U1636-536 (Zhang \etal 1996, fig. 2) for which 
the {\it rms} amplitude increases almost monotonically with the energy up 
to at least 20 keV.

\section{Theory and observations of kHz QPOs}

\subsection{Main observational results}

In this Section we shall summarize the main observational results on 
the kHz QPOs that need to be understood. 

The launch of RXTE opened up a new era in the study of QPOs. Recently,
the kHz QPOs have been discovered in the persistent fluxes of 8 LMXBs: Sco
X-1 (\cite{vdk96a}), 4U1728-34 (\cite{str96a}), 4U 1608-52
(\cite{vpar96}, \cite{ber96}), 4U 1636-53 (\cite{zha96}), 4U 0614+091
(\cite{for96}), 4U 1735-44 (\cite{W96}), 4U 1820-30 (\cite{sm96}), and
GX5-1 (\cite{vdk96b}).  In addition, episodic and nearly coherent
oscillations have been discovered during several type-I X-ray bursts 
from KS 1731-260 with a frequency of 363 Hz (\cite{str96a}), during
one type-I burst from KS 1731-260 with a frequency of 524 Hz
(\cite{morg96}), during 3 bursts from the vicinity of GRO J1744-28 with a
frequency of 589 Hz (\cite{slj96}), and during 4 type-I bursts from 4U
1636-53 with a frequency of 581 Hz (\cite{zha97}). The presence of a
pair of the kHz QPOs is characteristic of almost all these observations.
The centroid frequency of these QPOs ranges from 400 (4U 0614+091,
\cite{for96}) to 1171 Hz (4U 1636-53, \cite{vdk96c}). According to van
der Klis \etal (1997) the observations of Sco X-1 also show the
presence of additional two peaks at about 40 and 90 Hz simultaneously
with the kHz QPO.

For Sco X-1 (\cite{vdk96c}), the difference in the centroid
frequencies of two QPO peaks changes with time, whereas for 4U 1728-34
and 4U 0614+091 this difference is constant with time and does not
depend on the count rate (\cite{str96a} and \cite{for96}). In
particular, for 4U 1728-34 this difference is always 363 Hz, the same
as the frequency of nearly coherent oscillations observed during
several bursts. For 4U 0614+091, the difference in centroid
frequencies for the two QPO peaks coincides with the centroid frequency of
the third QPO peak which was observed during a 1/2-hour period.
In general, the centroid frequencies of such QPOs range from 400  
(4U 0614+091, \cite{for96}) to 1171 Hz (4U 1636-53, \cite{vdk96c}).

There are many similarities between the QPOs observed in different 
sources, and the most important common features are the following:

\begin{enumerate}

\item For some sources (\cite{for97}) there is a 
correlation between the QPO centroid frequency and the count rate.
However, Zhang  \etal 1998 found that the frequency-count rate correlation
is more complex in the case of Aql X-1 source.
This correlation persists over a short time scale, from minutes to
hours, and it apparently breaks dawn on longer time
scale in this source (which has also been observed in other sources, 
e.g. in 4U 1608-52 [Berger \etal 1996]). However, in 4U 0614+091 
(\cite{for97}) the correlation is rather good over a long time scale.    

\item Large Q-values, up to $\sim 10^2$. [There is also an indication that
the higher frequency QPOs become more coherent as the frequency and the
total count rate increase (\cite{vdk97}).]

\item The {\it rms} amplitudes of QPOs range from the low values (at the
threshold of detectability) to a maximum of 12 \% in the RXTE/PCA band
(2-60 keV). [For every source, where the data are available, the
{\it rms} amplitudes show strong dependence on energy.  For example,
for 4U 1636-53 the {\it rms} amplitude at 3 keV is only 4\%, while at
20 keV it is as high as 16\% (\cite{zha96}). Furthermore, the variable
hard X-ray flux seems to anticorrelate with the soft X-ray count rate
and with a value of the highest QPO frequency (\cite{for96}).]

\item In Sco X-1, there is a good correlation between the frequencies
of the 6 Hz, 40 Hz, 80 Hz and the kHz QPO. 
Separation of two peaks  anticorrelates 
with the kHz frequency and accretion rate. (see \cite{vdk97}).
\item The highest observed QPO frequencies 
fall in a remarkably narrow range from 1066 to 1171 Hz (\cite{zss97}).
 
\end{enumerate}

\subsection{Interpretation of observational results}

We suggest that a set of peaks seen in the kHz QPOs is essentially due to
the effect of rotational splitting of the main oscillation 
frequency (see Section 2.4). This effect should be very important for 
the accretion disk since the characteristic frequency of the gravitational 
oscillations of the CB region is of order of the Keplerian frequency.

To estimate the split frequencies, we adopt that 
$\Omega_0 \approx \Omega $, which is a good approximation for a 
nearly Keplerian accretion disk. Then, for the lowest-order modes 
with $\rm m = 0,~-1$ and k = 1, 2, 3, 4 and 5, the oscillation
frequencies are: $\rm \Omega_{k,0} \approx \Omega_0$, 
$\rm \Omega_{1,-1} \approx 2\Omega_0 /(\pi^2 s +2)$,
$\rm \Omega_{2,-1} \approx 2\Omega_0 /(4\pi^2 s +2)$, 
$\rm \Omega_{3,-1}\approx 2\Omega_0 /(9\pi^2 s +2)$, 
$\rm \Omega_{4,-1} \approx 2\Omega_0/(16\pi^2 s +2)$, and 
$\rm \Omega_{5,-1} \approx 2\Omega_0 /(25\pi^2 s +2)$,
respectively. For $m=-2$ and $k=1$ we have $\rm |\Omega_{1,-2}| 
\approx \Omega_0 [1-4/(\pi^2s+5)] $.

If we take $\rm \nu _0 \equiv \Omega _0/2\pi = 1200$ Hz and assume
that $\rm s \approx 0.7$, we get the following frequencies (arranged
in ascending order): $\nu _{5,-1} \approx 14$, $\nu _{4,-1} \approx 20$,
$\nu _{3,-1} \approx 40$, $\nu _{2,-1} \approx 80$, $\nu _{1,-1}
\approx 270$, and $\nu _{1,-2} \approx 800$ Hz. These frequencies match 
the observed QPO frequencies in LMXBs. It is important that for a given
mode with $\rm k\geq 2$ the ratio between frequencies is practically
independent of the function s and is solely determined by the
quantum numbers. As has been pointed out by  Hugh Van Horn (1997) 
the asymmetry between modes with $\rm m <0$ (prograde modes) and 
modes with $\rm m > 0$ (retrograde modes) in our interpretation of QPOs  
in terms of the rotational splitting may be akin to the result 
obtained by Carrol \& Hansen (1982) that for the nonradial stellar 
oscillations the slow rotation enhances the stability of retrograde 
modes and makes prograde modes less stable. Note also, that the 
contribution of a $\rm (k,m)-$component to the power spectrum 
is determined by a smoothness of a function over the corresponding 
coordinates. It is well-known from the Fourier analysis 
that the spectral power of the $l$-component scales as $l^{\rm -n}$, 
where n is an order of the highest existing derivative of 
a perturbation with respect to the corresponding coordinate 
(here $l= $m for azimuthal components, and 
$l= $k for $\rm z$-components). Thus, it is
natural to expect the presence of only lowest-order azimuthal
components ($\rm |m|=0,1,2 $) in the power spectrum, since it is 
very likely that the perturbations along the $\varphi-$coordinate 
are rather smooth. On the contrary, the vertical perturbations 
should be essentially discontinuous, and their Fourier spectra should 
contain a large number of components with $k > 1$ whose contribution 
to the power spectrum $\sim k^{-2}$. It is also worth noting that the 
spectral power of a $\rm (k,m)-$component is related with the spectral
power of a disk at the frequency corresponding to this particular 
component. It is demonstrated (Lyubarskij 1996, Kazanas, Hua \& 
Titarchuk 1997)  that the density of the power spectrum for  
a Shakura-Sunyaev disk behaves as $\omega^{-1}$ at the low frequencies.
Thus, because of this dependence, we can expect more power in  
the QPOs at lower frequencies. 

Therefore, we suggest that {\bf the rotational splitting of the 
main oscillation frequency of the CB region in 
the accretion disk may be responsible for the kHz QPOs in
LMXBs. The observed three frequencies, ranging from 300 to 1170 Hz,
can be naturally interpreted in the following way: the lowest 
frequency of 200-300 Hz corresponds to a mode with m = -1 and 
k = 1; the higher frequency of 800 Hz corresponds to a 
mode with m = -2 and k = 1, and the highest frequency 
is the main frequency of the oscillations.}

The main observational consequences of a CB model are:

\begin{enumerate}

\item Correlation between the total count rate and the QPO frequency
(see Figure 3).  The latter increases with an increase in the 
accretion rate or, equivalently, with an increase in the 
total count rate (cf. Section 3.1, item 1) provided 
that the viscosity is determined
by the turbulent motions. In the regime where the radiative viscosity 
dominates the frequency-count rate dependence may be more complex, and
the kHz QPOs may even disappear. Zhang \etal (1998) show that 
in the source Aql X-1 the total X-ray flux (the total 
accretion rate) does not seem to be the only factor that controls 
the frequency of the kHz QPOs.
{\it We argue, that this frequency-count rate dependence 
is rather determined by Reynolds number ($\gamma-$parameter). In other
words, it is the ratio of the accretion rate to the viscosity that determines 
the frequency-count rate behaviour for the kHz QPOs}.

\item Correlation between the Q-value and the total X-ray 
flux. When the illumination of the CB region (the accretion rate) 
increases, the temperature of the CB region should decrease
(see Figure 4 and equations [18], [24]).  

\item Observed spectrum is a result of Comptonization of soft 
radiation from a disk in the CB region (ST80). The kHz oscillations 
is a manifestation of the mechanical oscillations of the hot CB region
in the radial and vertical directions.

\item Partial oscillation amplitudes should be larger at higher
energies [exactly this correlation has been observed by \cite{zha96}
(see Section 3.1 for details)]. The QPO frequency should also
correlate with the total count rate and anticorrelate with the
hardness of a spectrum (see Fig. 4, and also \cite{for96}).

\item The CB effect should be more pronounced for the higher Reynolds 
numbers ($\gamma-$ parameters) for which the transition layer is closer 
to a NS surface, and for which the angular velocity in a disk $\Omega$ and
therefore the characteristic frequency of gravitational oscillations
of the CB region $\Omega_0/2\pi$ are larger.

\item Spacing between the QPO frequencies and the main oscillation
frequency (the highest QPO frequency) would decrease with increasing
in the accretion rate (or increase of the main oscillation frequency).
In fact the parameter $s$ (see Eq. A8) monotonically increases from 0 to
1 when the CB radius (or the oscillation) increases.   
The rotational splitting modes with the frequencies, 6 Hz, 40 Hz, 80 Hz,
have to correlate with the main frequency 1200 Hz (see Eq. A9).

\item The QPO oscillations are likely to become highly incoherent when 
the oscillating region plunges into the magnetopause, where the plasma 
turbulence enhances the viscosity and tends to suppress the
oscillations. In this case the QPO frequency should be limited by 
the Keplerian frequency corresponding to the Alfven radius, 
$\rm R_A\approx 1.4\times 
10^6 (B/4\cdot10^8~G)^{-0.8}\dot M_{18}^{2/5}m^{-1/5}$ cm, where $B$
is the magnetic field strength at the NS surface, 
$\dot M_{18}=\dot M/(10^{18}$ g s$^{-1}$) and $m=M/M_{\odot}$. Note
that the Alfven radius and 
{\it hence the highest frequency} only slightly depend on 
the accretion rate. If the masses and values of the magnetic field 
strength for NSs in bursters are 
within a very narrow range, then one can expect that the highest frequencies
for all these sources will also fall into a rather narrow range.  

\item
Morgan, Remillard \& Greiner  (1997) reported the detection of 67 Hz QPO from 
 GRS 1915+105 by RXTE.
They found the energy dependence of the 67 Hz QPO in  
their May 5 1996 observation, where the $rms$ amplitude increased with 
the photon energy from a level of 1.5\% below 
5 keV to 6\% above 13 keV. This correlation indicates that the 67 Hz QPO 
is associated with the most energetic spectral components visible
in the PCA (Proportional Counter Array on board RXTE).
In terms of our model it can be interpreted as the oscillation of 
the inner edge of the accretion disk (CB region), 
or as the $g-$mode disk oscillation in the region of radius 
of 4 $r_s$ (see for details Appendix B). Such a correlation should be 
more pronounced in the model accounting of the flux variation in the 
converging-inflow produced hard tail (Titarchuk, Mastichiadis 
\& Kylafis 1997, Titarchuk \& Zannias, 1998), because of substantial 
variation in the exposure of the converging inflow region (the 
characteristic radius of which is $2r_s$) to the soft 
seed photons from the disk.   
\par
\noindent   
The similar energy dependence is observed for 300 Hz QPO detected
by RXTE (Remillard \etal 1996) from  another superluminal source, 
GROJ1655-40. And thus we can argue that {\it the QPOs observed from BH system 
detected only in the high energy bands reveal the scale of the high 
energy emission region  and hence  the nature of the high energy formation
in the soft states of BH systems in terms of the converging inflow 
Comptonization}.

\end{enumerate}

%In conclusion of this Section it must be pointed out that for the BH
%systems all the abovementioned effects and correlations should hold,
%with an appropriate rescaling to the lower QPO frequencies.

\section{Alternative models}

\subsection{Acoustic waves}

The acoustic waves standing in the hot area surrounding a NS is an
alternative possibility for the interpretation of the kHz QPOs. The
standing shocks above a NS surface can serve as boundaries for such
waves. The existing controversy about the stability of such shocks 
will be probably resolved by 2-D simulations (\cite{chen97}) 
indicating that after an initial transition phase of rapid shock 
movements a system relaxes to a stable configuration comprising the 
quasi-stable, hot torus around a NS. These simulations show that 
after passing a shock, the accreting gas becomes involved in the 
complex, predominantly tangential vortex motions with a very small 
radial Mach number in the postshock region. The resulting picture 
is quite distinct from a simple, quasi-virial spherical accretion: 
it allows for the existence of a hot coronal postshock region around 
a NS perfectly suitable for a sound wave propagation. The acoustic 
oscillations of such a region can manifest themselves as the kHz
QPOs. Note that the NS systems seem to be more favorable, 
since the standing acoustic waves may oscillate between two 
surfaces, the stellar surface and the outer boundary. Moreover, 
these surfaces can reflect, absorb and even emit sound waves.

The effects of an interaction of the shock surfaces with the sound
waves are extensively discussed by Landau \& Lifshitz (1989). In
particular, the strong shocks are subject to corrugation instability
when a shock surface becomes rippled. In the range of angles between
the wave vector and a normal to the shock surface the reflection
coefficient for an incident sound wave may exceed unity or even
diverge. This means that the sound waves can be not only reflected but
can also be spontaneously emitted by a shock (into the postshock region), so
that a sound wave can receive energy from a shock. The damping of
sound waves by a shock can also occur under certain circumstances.  For
example, for a sound wave falling normally from a postshock side onto
a flat shock surface, the reflection coefficient is, in general, $<$
1. Note that almost 100\% reflection takes place when a gas is 
isothermal near the strong shock, and this may well be relevant to 
the physical situation under consideration. Due to Comptonization of 
the ionized gas the outer layers of a postshock region adjoining the 
shock are always isothermal. It is well-known that a similar 
situation occurs in the outer atmospheric layers of X-ray bursters, 
and in many other situations where the electrons in the outer, 
optically thin layers of gas clouds are heated by Comptonization and 
kept at a constant temperature equal to the spectral temperature of 
radiation emerging from a cloud. The ratio of specific heats 
is equal to unity and the reflection coefficient for a sound wave is 
almost equal to unity as well.

However, it is unlikely that all sound waves to be
perfectly reflected by a boundary of a hot region, regardless of its
physical nature. The accretion disk, for example, may contribute to
the wave damping, while the asymmetry may, as is mentioned above,
result in a spontaneous emission and, therefore, in an enhancement of a
sound wave. It is important that the standing acoustic waves in a 
postshock region is essentially a resonance response of a dynamical
system (the lobe) to the background noise, which is always present in
a postshock region and contains all possible frequencies.  The balance
between a continuous excitation of resonant modes and dissipation of
sound waves determines their finite amplitudes.

The waves under discussion are usual sound waves propagating in a fully
ionized gas. The protons and electrons are coupled together by strong
electrostatic forces, and the force driving the oscillations is a
gradient of thermal pressure in an electron-proton mixture. In the
situation characteristic of the inner boundary of an accretion disk
the proton temperature, $\rm T_p \sim$ few keV is of order of the
electron temperature, $\rm T_e$.  When entering a
postshock region the protons instantly get much higher temperature, of order of
the virial one, $\rm T_v \sim$ few MeV. However, because of frequent
collisions with much cooler electrons the protons lose their energy
very fast, over a timescale $\rm t_{p, cool} \approx 5.8 \, T_e^{3/2}
(M_p/m_e)^{1/2}/ (n_e \ln \Lambda)$ s (Spitzer 1956). Given $\rm T_e
\sim 10 \,$ keV, $\rm n_e \sim 10^{17}\, cm^{-3}$, and $\rm \ln
\Lambda \approx 13$, we get $\rm t_{p,cool} \approx 2.5 \times
10^{-4}\, s$.  The sound speed is therefore determined mostly by
protons $\rm c_s=(\partial p/ \partial \rho)^{1/2} \approx 0.98
\times 10^8 T_{5}^{1/2}$ cm s$^{-1}$. Since protons and electrons are coupled
by electrostatic forces, the frequent Coulomb collisions of protons
determine the damping of sound oscillations, and we can use 
the proton mean-free path, $\rm {\it l}_p$, in the standard expression 
for the wave damping in a fluid. The decay coefficient is  
$\rm \gamma_{d}~=~ |\dot E_{m}|/E $, where $\rm \dot E_{m}$ is 
an average rate of acoustic energy dissipation, and E is a total 
acoustic wave energy.  Following Landau \& Lifshitz (1989), we can
write for the decay coefficient $\rm \gamma_d\approx 
(2\pi\nu)^2({\it l}_p/c_s)$, where $\nu$ is a sound frequency.

Since the damping of sound oscillations in a hot plasma is mostly due to
Coulomb collisions, the value $\rm {\it l}_p$ can be estimated as $\rm
{\it l}_p=(H/\tau_0)(\sigma_T/\sigma_C)$, where H and $\tau_0$ are the
characteristic length-scale and the Thomson optical thickness of the
plasma, respectively. The ratio of Coulomb to Thomson cross-sections 
is $\rm \sigma_C/\sigma_T\approx 2.35\cdot 10^5 T_5^{-2}$, where $\rm
T_5=T/5$ keV.  The characteristic length-scale of an atmosphere, H, 
can be expressed in terms of the observed sound wave frequency as 
$$
\rm H = (\pi/ 2)(c_s/2\pi \nu) = 2.4\cdot T_{5}^{1/2}\nu_{kHz}^{-1}
\cdot 10^4~~~{\rm cm},\eqno(34)
$$
where $\rm \nu_{kHz}=(\nu/1~kHz)$. Assuming that H=const, we get 
$\rm \nu \propto T^{1/2}$. The proton mean free path can be estimated  
(by using the above relations between $\rm {\it l}_p$ and H) as 
$\rm {\it l}_p = 0.1 \cdot T_5^{2.5}\nu_{kHz}^{-1}\tau_0^{-1} ~{\rm cm}$.
Thus, one can find that 
$\rm \gamma_{d}\sim 4\cdot 10^{-2}\nu_{kHz}T_5^2\tau_0^{-1} \, s^{-1} $,
and the characteristic decay time  
$\rm t_\ast=1/\gamma_{d}=25\tau_0\nu_{kHz}^{-1} T_5^{-2}$  ${\rm s}$. 
This means that at high
temperatures the oscillations should decay very fast.  Note also that
the oscillation frequency may vary as a result of a changing of the
boundary condition at the shock. For example, in the case of a free
boundary the oscillation frequency is a factor of 2 higher than that 
for a fixed boundary (the estimated size of a shock is a factor of 2
larger than the characteristic length scale given by [34]). The 
radiation pressure and/or viscosity can also play an
important role in damping of sound waves, despite the fact that the
radiation pressure is small compared to a gas pressure at luminosities
well below the Eddington limit (see \S 2.2).

In the low (hard) state of a NS system the temperature of a 
postshock region is higher, $\sim$ few $\times 10$ keV, and the acoustic
waves die off rapidly. They can, therefore,
exist only in the high (soft) state where $\rm T_e \sim$ few keV,
and the damping is small.

The plasma oscillations usually considered in the laboratory plasmas 
do not propagate under the conditions we discuss here for a very simple
reason. The effects of collisionless plasma like plasma waves, Landau
damping of ion-acoustic waves, etc. do not apply to our situation
since the plasma we are dealing with is essentially collisional. 
Given a short proton mean-free path, $\sim 0.1$ cm, and high 
velocity, $\rm {\it v}_p \sim 7 \times 10^7 $, cm s$^{-1}$,
the proton collision frequency is $\rm \nu_{col} \sim 7 \times 10^8$
Hz that exceeds a frequency of a kHz wave.

As far as the observed amplitude of flux oscillations is concerned, it
can be interpreted similarly to a CB model (Section 2), in terms of
a variable area of an emitting surface.

\subsection{Rotation of a NS and oscillations of a disk}

Let us briefly address the issue of whether a NS itself may be a source
of e.g. 363 Hz nearly coherent oscillations (S96). In general, 
in an oscillating
system, the amplitudes of any two modes $\rm {\bf A}_1 
e^{i \omega_1 t}$ and $\rm {\bf A}_2 e^{i\omega_2t}$, may be either 
added to or multiplied by each other. When the excitation mechanism 
for the modes is the same, the addition rule applies, and Fourier 
spectral analysis will reveal two independent peaks in the power 
density spectrum (PDS), at $\omega_1$ and $\omega_2$. 
This is the case, for instance, for a set of overtones of the same
physical origin differing from each other by the mode quantum numbers 
only. When two oscillatory motions are of different origin, then 
we should multiply their amplitudes. The ``beating'' motion 
belongs to this category, including the ``beat-frequency'' model for
the low-frequency QPOs in LMXBs (see Lewin, Van Paradijs, \& Taam 1993
for a review). In this case both $(\omega_1+\omega_2)$ and
$\vert \omega_1-\omega_2 \vert$ peaks should be present in the PDS of 
a system unless $\omega _1 = \omega _2$. 

In LMXBs a NS is thought to be weakly magnetized and the lack 
of pulsations even in quite bright LMXBs apparently supports this
idea. It is quite reasonable to assume that the nearly coherent 
$\sim $363 Hz oscillations are associated with the stellar rotation: 
a relatively high 
temporal stability of a frequency and its independence of the 
spectrum and flux are the main arguments in favor of such a
possibility. The 363 Hz oscillations occurred only during type-I 
X-ray bursts from 4U1728-34 (other 3 LMXBs did not show any type-I 
bursts during the observations). The explanation this QPO in terms 
of a NS rotation encounters a difficulty that the
emitting region should be seen only during a very short time (during
bursts), which is, in principle, possible e.g. in a model of
inhomogeneous combustion on a NS surface (see Bildsten 1995).
The accretion disk trapped $g$-modes is another possible
mechanism that may be responsible for the stable $\omega_3$-oscillations
(see Appendix B). Similar to a NS rotation, these oscillations, when 
superimposed on the intrinsic oscillations of a hot region around
a NS may produce multiple QPO peaks. For the type-I bursts with a 
relatively short recurrence time (high values of $\rm \dot{M}$) 
the surface layers of a NS, remain convectively stable, and a 
combustion front propagates over the stellar surface very slowly, 
at $\rm v_{comb} \sim 300$ cm s$^{-1}$ (Bildsten 1995). This means 
that a combustion inhomogeneity (e.g. associated with a more active 
burning in one zone where it started first) can persist for a long 
time, at least $\sim 10^4$ s, provided that there is enough fuel. 
Thus, on a timescale of $\sim 15$ s (typical duration of a type-I 
burst) a combustion inhomogeneity can manifest itself as a 
rather stable pattern of energy release on the stellar surface, 
which seems to agree with the 4U1728-34 data. The narrow PDS peak 
implies a high Q-value, again in agreement with the observations. 

The period evolution of nearly coherent 363 Hz oscillations in 
4U1728-34 during a burst can be understood in terms of 
the hydrodynamics of a "hot spot" produced by an accreted gas 
on the stellar surface (in a boundary layer). For
example, as it has been suggested by Fujimoto (1988), the spreading
of the accreted matter over the surface of a weakly magnetized NS 
accreting from a disk should be accompanied by a differential rotation
in the very surface layers. This (radial) differential rotation gives rise 
to the hydrodynamic instabilities which will redistribute 
(through the development of a turbulence) the angular momentum in 
the surface layers of a NS. For the physical conditions characteristic
of the surface layers of an accreting rapidly rotating NS 
the baroclinic instability (see Pedlosky 1979, Knobloch \& Spruit
1982, Zahn 1983, and Fujimoto 1987 for review) is perhaps the 
most important one. This instability is related to the fact that in a 
differentially rotating star in hydrostatic equilibrium the surfaces 
of constant density in general do not coincide with the surfaces of 
constant pressure. The growth rate for this instability can be
estimated as (see e.g. Fujimoto 1988) $\rm \tau _{bc}^{-1} 
\lsim 0.6 \Omega _0 \cos \theta _0$, where $\Omega _0$ is the local 
angular velocity at the point with a colatitude $\theta _0$ ($< \pi /2$). 
This estimate shows that the instability may develop as shortly as 
during 1-10 stellar revolutions. The development of the 
hydrodynamic instabilities means the occurrence of the largescale plasma 
flows and mixing of a matter in the surface layers of a NS that may 
eventually destroy a coherency of e.g. 363 Hz oscillations. 

Two other mechanisms have been proposed in the literature to explain
the kHz QPOs. Recently, Klein \etal
(1996) have proposed that these QPOs result from a turbulence occurring
in the settling mounds of the neutron star polar caps. Their numerical
simulations produce QPOs in the kHz range with fractional $rms$ 
amplitudes of order of 1 \%. Several QPO frequencies appear to be
present in their simulations, with the highest ones are above 2 kHz.
Miller, Lamb and Psaltis (1997) have proposed a model in which the QPO with
higher frequency is the Keplerian frequency at the sonic point 
beyond which the radial inflow velocity becomes supersonic. 
The position of the sonic point is determined by the radiation forces
which remove the angular momentum from the accreting matter. In this
model the accretion flow is also modulated by the radiation flux from 
a NS with its spin frequency. This accretion-flow
modulation causes the Keplerian frequency to beat with a NS spin frequency
to produce QPO with a lower frequency. In this model, there is an 
upper limit for the QPO frequencies, either the Keplerian frequency at the
stellar surface, or the frequency of the marginally stable orbit.

We must admit that the variety of
observational phenomena, such as e.g., the twin QPO
peaks with the constant and (in some cases) variable  difference between them, 
as well as appearance
of a separate peak at the frequency corresponding to this difference
need to be understood. 
We believe, the models of CB oscillations we attempted to address in this paper,
may  provide a reasonable 
and consistent interpretation for the most experimental facts
related to the phenomena of kHz QPOs, both for the NS and BH binary
systems.

It is quite possible that $\sim 200 - 400$ Hz PDS peaks manifest
either the proper spin frequency of a NS, or $g$-mode 
oscillations in an accretion disk.  The $g$-modes can be observed at 
frequencies of $\sim 200 - 500$  and $\sim (60 - 300) \times 
(5/m)$ Hz for the NS and BH systems, respectively. The 
disadvantage of $g$-modes from the observational point of view is 
that they should be localized in a rather small region of a disk 
with $\rm \Delta R\sim GM/c^2$. 

The crucial test of potential observability of $g$-modes in a disk 
around a relativistic compact object may be provided by the millisecond 
timing of BH transients in the high (soft) state where the luminosity is
dominated by the thermal radiation from the accretion disk. 
Such $g$-modes should have frequencies $\sim 200\cdot (5/m)$ Hz. 
The QPOs with the frequencies $\sim 67$ and $\sim 300$ Hz   
were recently detected by RXTE from GRS 1915+105 (Morgan \etal 1997)
and from GROJ1655-40 (Remillard \etal 1996), respectively, 
during the very high states of these sources. These observations 
may suggest the occurrence of $g$-mode disk oscillations, 
at least in these systems  (see, also discussion in \S 3.2).
The $g-$modes can be observationally distinguished from the CB
oscillations due to the fact that the $g$-mode eigenfunctions are 
symmetric with respect to some radius in the disk (Nowak \& Wagoner
1992), and thus the effect of rotational 
splitting will be smeared out for these modes (see Eq. A1), which 
is certainly not the case for the CB oscillations.    

\section{Conclusion}

We propose a model that can self-consistently explain most of the
observational facts on kHz QPOs. The main physical ingredients of our
model are:

\begin{enumerate}

\item The possibility of a super-Keplerian rotation (centrifugal
barrier, CB) in a boundary region between a Keplerian accretion disk
and a compact object. The formation of kinks, weak discontinuities 
in a supersonic accretion flow, makes an adjustment of the disk
rotation to the innermost boundary (NS either, magnetosphere surface  
or the last stable orbit around a BH) to occur through the 
dynamical episodes of a locally super-Keplerian rotation.

\item This dynamical adjustment is dictated by the turbulence of flow,
Reynolds number, i.e.  the ratio of the mass accretion and 
viscosity dictates the regime of the QPO behaviour. 

\item The excitation of relaxation nonradial oscillations in the CB
region.

\item The effect of rotational splitting of the main oscillation
frequency (of order of the local Keplerian frequency) of the CB
region. 

\end{enumerate}

We have demonstrated that the CB effect can explain various 
correlations and anticorrelations found in the observations of 
kHz QPOs (see \S 3.2 for details). Also, we have shown that the 
splitting of the main kHz frequency due to a disk rotation 
produces a discrete spectrum of frequencies that perfectly match 
the observed QPO frequencies (see \S\S 2.4 and 3.2 for details). 
In our model the QPO frequencies are inversely proportional to 
the mass of a compact object. We have therefore suggested that 
a similar phenomenon should be observed in the BH systems too, 
with the QPO frequencies should be a factor of 5-10 lower than in the
case of the NS systems. In addition, in our model a typical size of 
the emission region is estimated to be $\sim 1-3$ km, in a good 
agreement with the time-lag measurements. We have briefly discussed 
some alternative models that may be relevant to the physics of QPOs 
including the standing acoustic waves in the postshock region around a
NS, the effect of frequency modulation due to the proper rotation of 
a NS, and $g$-mode intrinsic oscillations of an accretion disk.

The authors thank NASA for support under grants NAS-5-32484 and 
RXTE Guest Observing Program. I.L. thanks the Isaac Newton Trust
(Cambridge) and PPARC for support. The authors acknowledge discussions 
with Hugh Van Horn, Martin Rees, Guy Miller, 
Tod Strohmayer, Jean Swank, Nick White, Will Zhang. 
 Particularly, we are grateful the anonymous referee 
whose  constructive comments and criticism significantly improved this paper.

\section*{APPENDIX A. ROTATIONAL SPLITTING EFFECT}

In this Section we estimate the effect of a disk rotation on the
QPO frequency. The detailed theory for a slowly rotating star is given by
Ledoux \& Walraven (1958) and Unno \etal (1979). Here we calculate the
splitting of the eigenfrequency due to Coriolis force in the case of a 
disk geometry.

It must be pointed out that in our analysis the correction to the main
frequency $\rm \sigma ^{(1)}$ produced by Coriolis force is always 
smaller than the main frequency $\sigma _0$ (even if the main
oscillation frequency itself is comparable to the rotation
frequency). Thus, to derive this correction, we are still allowed 
to exploit the linear approximation and use the unperturbed values 
for the displacement vectors. [The fundamental difficulty in the 
problem of nonradial oscillations in the presence of rotation is 
that the hydrodynamic equations are not t-invariant, as the Coriolis 
term is replaced by the corresponding term with the minus sign 
by time-inversion. This means that the eigenvalue problem differs 
from its complex conjugate, or, that the displacements 
$\pmb{$\xi $}$ and $\pmb{$\xi $}^{\ast }$ are not eigenvectors
of the same eigenvalue. That is why the effect of Coriolis force 
on the nonradial oscillations can be taken into account only for 
the case of a slow rotation for which $\pmb{$\xi $}$ and 
$\pmb{$\xi $}^{\ast }$ can be approximated by the corresponding
vectors determined from the eigenvalue problem without rotation, 
$\pmb{$\xi $}^{(0)}$ and $\pmb{$\xi $}^{(0)\ast }$), respectively.]

For the axisymmetric case (cf. Unno \etal 1979, 
equation [18.27]) the correction to the eigenfrequency reads
$$
\rm \sigma^{(1)}={{-i \int_{D_{cb}}[{\pmb{$\Omega $}} \times 
{\pmb{$\xi $}}^{(0)}_{k,m}] {\pmb{$\xi $}}^{(0)\ast}_{k,m} dD_{cb}}
\over{\int_{D_{cb}}{\pmb{$\xi $}}^{(0)}_{k,m} 
{\pmb{$\xi $}}^{(0)\ast}_{k,m} dD_{cb}}},
\eqno(A1)
$$ 
where $\rm {\pmb{$\xi $}}^{(0)}_{k,m}(t,r,\varphi, z)$ is the $\rm
(k,m)$-displacementent component. For a disk- (ring-) like geometry
the displacement component should be calculated by using the $\rm
(k,m)$-harmonics of a complete set of eigenfunctions for a  
disk- (ring-) like configuration, $\rm \{u_{k,m}\}$. In equation (A1) 
the integration has
to be taken over the disk area $\rm D_{cb}$ of radius R, width $\rm
\Delta R$, and half-thickness $H_d$.  The appropriate displacement
component $\rm {\pmb{$\xi $}}^{(0)}_{k,m}$ can be written in cylindrical
coordinates as (see e.g. Unno \etal 1979, equation [18.28])
$$
\rm {\pmb{$\xi $}}^{(0)}_{k,m}=\left[\xi_r,~ \xi_{\varphi} {{\partial} 
\over{\partial \varphi}},~ \xi_z r {{\partial}\over{\partial z}} 
\right] u_{km},
\eqno(A2)
$$ 
where the functions $\rm u_{k,m}$ satisfy 
the  free-boundary conditions at $\rm z=H_d$ and  $\rm z=-H_d$ (see e.g. Morse
\& Feshbach 1953), and the symmetry (with respect to the disk plane) 
condition. These conditions are 
$$
\rm {{\partial u_{k,m}}\over {\partial z}}(t,\varphi,H_d)
={{\partial u_{k,m}}\over {\partial z}}(t,\varphi,-H_d)=0,
\eqno(A3)
$$ 
and
$$
\rm u_{k,m}(t,\varphi,z)=u_{k,m} (t,\varphi,-z), 
\eqno(A4)
$$
respectively, where $\rm 0\leq z\leq H_d$. We must note that the 
symmetry condition is not warranted in the case of a disk tilt 
above the plane $\rm z=0$ on one side and below the plane on 
the other (Van Horn 1997).

In the case of uniform rotation, the temporal and azimuthal dependences 
of eigenfunctions $\rm u_{k,m}$ are taken as 
$\rm {\it e}^{i(m\varphi-\Omega t)}$ (Unno \etal 1979).  
Thus, the eigenfunction component $\rm u_{k,m}$ can be written as 
$$
\rm u_{k,m}={\it e}^{im\varphi}\cos(\pi kz/H_d){\it e}^{-i\Omega t},
\eqno(A5)
$$
where $\rm m=\pm 1,\pm 2,....$, $k=1,2,....$.

The integral $\rm I_1$ in the numerator of equation (A1) 
is given by
$$
\rm I_1=-2\pi H_dR\Delta R [2\Omega m \hat\xi_r\hat \xi_{\varphi}],
\eqno(A6)
$$
where $\rm \hat \xi_r$ and $\rm \hat \xi_{\varphi}$ are 
the average displacements over r- and $\varphi -$coordinates,
respectively, for a disk-(ring-) like configuration.
The integral $\rm I_2$ in the dominator of equation (A1) is
$$
\rm I_2=2\pi H_dR\Delta R \left[\hat \xi_r^2+(\hat \xi_{z} R/H_d)^2\pi^2k^2+ 
m^2\hat\xi_{\varphi}^2\right],
\eqno(A7)
$$
where $\rm \hat \xi_{z}$ is the average displacement over z-coordinate.
Here $\hat \xi_r$, $\hat \xi_{\varphi}$, and $\rm \hat \xi_{z}$ are
the average displacements for a disk configuration in radial,
azimuthal, and vertical directions, respectively. To illustrate the
effect of rotational splitting we may justifiably assume that 
$\rm \hat\xi_r\sim\hat \xi_{\varphi}$, then  
we arrive at
$$
\rm \sigma^{(1)}=- {{2m\Omega}\over{1+s\pi^2k^2+m^2}},
\eqno(A8)
$$ 
where $\rm  s(R/H_d)=(\hat\xi_{z}R/\hat\xi_{r}H_d)^2$ is a function
determined by the vertical structure of a disk. For more or less 
realistic structure of a disk $\rm s\lsim 1$ and depends on the ratio 
$\rm R/H_d$, which is either $\gsim 1$ or $\gg 1$. 

Thus, the frequency of the oscillations as seen by a distant observer 
is given by (cf. \cite{unn79}, equation [18.33])

$$
\rm \Omega_{k,m}= \Omega_0+m\left[1- {{2}\over{1+s\pi^2k^2+m^2}}\right]\Omega. 
\eqno(A9)
$$ 

In conclusion, we must note that this formula can also be obtained 
as follows. If the relative correction to the main frequency produced
by the Coriolis term is small, then in the equation of motion for the 
perturbation we can use the displacement vector corresponding to the 
eigenvalue problem without rotation ${\pmb{$\xi $}}^{(0)}$. For a given
oscillation mode the right-hand side of equation of motion (containing
the gradient of a perturbed gravitational potential and the gradient 
of a perturbed gas pressure divided by density) can be replaced by 
the $\rm - \sigma _0^2 {\pmb{$\xi $}}^{(0)}$. Then, 
by multiplying both sides of the equation of motion 
by $\rm {\pmb{$\xi $}}^{(0)\ast }$ and integrating over the volume,
we get the following quadratic equation $\rm \sigma '^2 - 2 \sigma '
\sigma ^{(1)} - \sigma _0^2 = 0 $, where $\sigma '$ is the oscillation
frequency in the corotating frame, and $\sigma ^{(1)}$ is the
correction to the main frequency due to the Coriolis term. The
solution of this equation is $\rm \sigma ' = \sigma ^{(1)} + 
\sigma _0 [1+(\sigma ^{(1)}/\sigma _0)^2]^{1/2}\approx \sigma _0 + 
\sigma ^{(1)}$ (if $\sigma ^{(1)}/\sigma _0 \ll 1$), which leads to
the formula identical to (A9). Here we have ignored a solution with
the minus sign, because it gives very high frequencies.

\section*{APPENDIX B. $g$-MODE DISK OSCILLATIONS}

Here we calculate the characteristic oscillation frequencies for the trapped
g-modes in an accretion disk exploiting the results obtained by Nowak \&
Wagoner 1991, 1992, and 1993 (hereafter NW1, NW2, and NW3,
respectively). The two types of oscillations that are intrinsic to the
accretion disks around the compact relativistic objects and are of
potential interest for our present study  are $p$-
and $g$-modes.  The $p$-modes are longitudinal oscillations
propagating along the disk and having displacements in the radial 
direction (in a disk plane). It is very unlikely that these
oscillations can be observable.  The $g$-modes are mostly transverse
oscillations, and the gravity is the main restoring force for them. 
Their wave vector is always in the plane of a disk, but the
displacements are mostly in the perpendicular, $z$-direction.

The essential feature of both $p$- and $g$-modes is that they are 
trapped in a rather narrow range of radii. The possibility of the 
very existence of wave motions is essentially due to the
characteristic feature of the Keplerian law, the stability of orbits. 
In the Newtonian gravity the orbits are stable at any orbital
distance. The static general-relativistic effects e.g. in the case of 
Schwarzschild metrics result in that the stable orbits can exist only 
at $\rm R\geq 6GM/c^2$. For the modified Newtonian potential, 
$$
\rm \Phi = {{GM}\over R} \cdot \left[1-3{{GM}\over{Rc^2}}
+12\left({{GM}\over{Rc^2}}\right)^2\right] ,
\eqno(B1)
$$
the so-called 
epicyclic frequency (see
NW1) $\rm \omega _{e}=(c^3/GM)(r_m^{-3}-36r_m^{-5})^{1/2}$ vanishes at 
$\rm r_m=6$, thus reflecting the lack of stable orbits at $\rm r_m<6$.
Here $\rm r_m$ is the radial cylindrical coordinate in units of $\rm GM/c^2$,
and M is the mass of a central object.

The epicyclic frequency $\rm \omega_e$ has a maximum at $\rm r_{max}=
\sqrt{60}$. The $g$-modes are trapped near this maximum vanishing near
$\rm r_m\simeq 6$ and at $\rm r_m> r_{max}$. The characteristic width of a 
trapping region is therefore $\rm \Delta r_m \sim 1$, and the $g$-modes 
die off exponentially outside this region over characteristic length
scale $\sim $ a disk thickness, $h$. The trapping at
large radii is due to the pressure acting against the gravitational
restoring force. The elasticity coefficient in the restoring force is
$\rm \sim 1/R^3$, while the pressure is a much weaker function of R. 
As a result, the effective restoring force vanishes at 
$\rm r_m \simeq r_{max}+1\simeq 9$.

The equation governing the radial displacement $\rm \delta u$ as a 
function of r (NW2, equation [2.10]),
$$
\rm \omega^2c_s^2 (\delta u)_{rr}=
-(\omega^2-\gamma \Upsilon \Omega^2)(\omega^2-\omega_{epic}^2)\delta u ,
\eqno(B2)
$$
coupled to the corresponding equation for the oscillations in z-direction
(through a slowly varying with r function $\rm \Upsilon[r]$), fully
describes the main propagation properties of waves, including a  
characteristic radial extent of the trapping region, and effective
damping of waves out of this region.
\par
The dispersion relation for the waves describing both $p$- and
$g$-modes is 
$$
\rm \omega^2c_s^2 k_r^2 =
(\omega^2-\gamma \Upsilon \Omega^2)(\omega^2-\omega_{e}^2) ,
\eqno(B3)
$$
where $\Omega $ is a Keplerian frequency, and $\rm c_s$ is the sound
speed. The values $\rm k_r,~\Upsilon$ and $\Omega $ in the
dispersion relation are the values for the corresponding functions at
$\rm r_m\simeq r_{max}\simeq 8$ (NW2); $\rm \omega_{e}(r_{max}) \cong 0.029 c^3/
(GM)$, and $\Upsilon $ practically does not depend on r and varies 
from $\sim 1$ to $\simeq 12$ for the first five eigenmodes, 
depending on a mode number.

The $p$-modes correspond to $\rm \omega^2 > \gamma \Upsilon \Omega
^2$, $\rm \omega ^2 > \omega ^2_{e}$. As it has been pointed out in
NW3, the $p$-modes are trapped in a very narrow region, 
$\rm \Delta r_m \sim 0.1 GM/c^2$, and, since they have displacements 
in the disk plane, these modes are unlikely to be observed. On the 
contrary, the $g$-modes have frequencies below $\rm \omega_{e}$ and 
satisfy the conditions $\rm \omega^2<\omega^2_{e}$ and $\rm \omega ^2 < 
\gamma \Upsilon \Omega ^2$. Although the eigenvalue $\omega $ can 
be obtained from the exact solution of equation (B2) supplemented with
a corresponding equation for the z-component alone, we may reliably 
estimate it from the dispersion relation (B3). For this purpose, 
let us rewrite (B3) in the form
$$
\rm \omega^2=\omega^2_{e}+c_s^2k_r^2\cdot{{1}\over{1-\gamma\Upsilon\Omega^2/
\omega^2}}.
\eqno(B4)
$$
The frequencies $\rm \omega^2 < \omega ^2_{e}$ are allowed, and we can 
look for the solution of (B4) in the form $\rm \omega=\varepsilon
\cdot \omega _{e}$, where $\varepsilon \lax 1$. The hydrostatic 
equilibrium in the vertical direction implies that the sound  speed, 
$\rm c_s^2=\gamma P/\varrho=\gamma h^2\Omega^2$ (SS73), satisfies the 
relation $\rm c_s^2 k_r^2=\gamma h^2\Omega^2(2\pi/\lambda)^2$, 
where $\lambda $ is the radial wavelength approximately equal to 
the size ($\rm \simeq GM/c^2$) of a region with a mode trapping.
Note that $\rm \gamma\Upsilon [\Omega(r_{max})/ \omega_{e}(r_{max})]^2/ 
\varepsilon^2 \simeq (5/3)(1 \div 12)\cdot 2.25\cdot \varepsilon^{-2} 
\gg 1$, and equation (B4) yields 
$$
\rm \varepsilon^2\simeq \left[1+
\Upsilon^{-1}(2\pi h/\lambda)^2 \right]^{-1} .
\eqno(B5)
$$
For a reasonable assumption that $\rm h \sim\lambda(<1)$, and given a
typical value $\Upsilon \simeq 6$, we obtain $\varepsilon \simeq
0.4$. Thus, the value $\simeq \omega_{e}/2$ is a typical 
value for a circular frequency for one of the lowest-order 
$g$-modes. Given $\rm f_{e}=(\omega_{e}/2\pi)(1-2/r_{m})^
{1/2}=795/m~{\rm Hz} $ (here $\rm m=M/M_{\odot}$), these $g$-modes 
should have frequencies $\rm f=\omega/2\pi\simeq 200-500$ Hz.
These values perfectly fall in the range of frequencies between 
$\sim 250$ and $\sim 360$ Hz, observed from Sco X-1 and
4U1728-34, respectively.

\subsection*{Stability and damping of g-mode disk oscillations}

The Q-value for the $g$-mode oscillations is $\rm Q=f/\Delta f$, 
where $\rm \Delta f$ is determined by damping. The damping of $g$-modes 
is due to turbulent motions. The upward energy flux along the 
z-axis in $g$-modes due to turbulent motions can be estimated as 
(Stein 1967) $\rm F^z_{grav}\approx (20 \div 200)
({\it l}_{t}/H_d)^5\varrho u_0^3/{\it l}_{t}$. A numerical factor in 
this formula depends on the assumed spectrum of turbulence. The energy
density in a Keplerian disk is 
$\rm E=\varrho v^2_{\varphi}/2= (\varrho v_s^2/2)(R/H_d)^2 $, 
and the damping constant can be estimated as
$$
\rm \gamma_{d}={{F^z_{grav}}\over {E}}=(20 \div 200)
\left({{{\it l}_{t}}\over{H_d}}\right)^5
{{2\alpha^3 v_s}\over{{\it l}_{t}}}\left({H_d\over R}\right)^2. 
\eqno(B6)
$$
In derivation of this formula we have used the same relation between the
characteristic turbulent velocity, $\rm u_0$, and the sound speed,
$\rm v_s$, as for the  $\alpha $-disks (SS73): $\rm u_0=\alpha v_s$. 
Also, we assumed that the length scale of turbulent eddies 
$\rm {\it l}_t < H_d$. Taking into account that $\rm v_s/H_d=
v_{\varphi}/R=\Omega$ and $\rm \Omega(r_m) = 
9\cdot10^{3}\cdot m$ $s^{-1}$, we get $\rm \gamma_{d}=4.3\cdot 
10^{-3}\psi ~{\rm s}^{-1}$, where $\psi=m^{-1}(\alpha_2)^3
[({\it l}_{t}/H_d)_{1/3}]^4 (T_1)^{1/2}(H_d/R)^2$, 
and parameters $\rm \alpha_2$, $\rm {\it l}_t/H_d$, and 
$\rm T_1$ are normalized by $10^{-2}$, 1/3, and 1 keV, respectively.
The resulting estimate is Q$\rm \simeq 500~{\rm Hz}/\gamma_{d}
\sim 10^5\cdot \psi^{-1}$. The ``experimental'' Q-value for 
4U1728-34 can be estimated from the fact that the oscillations 
persist for a whole duration of the burst, 15 s, which gives 
$\rm Q_{exp}\geq 360~Hz/(1/15)=5.4\cdot 10^3.$

\subsection*{Luminosity variation produced by a disk $g$-mode 
oscillations}

To estimate the $g$-mode oscillation amplitude we can use
the temperature distribution in a disk calculated by SS73.
The relative contribution to the total luminosity, $\rm L_{tot}$, 
from the annulus in a disk confined by the radii from $r_m=$7.5 to 8.5, 
where the $g$-modes oscillations are trapped, can be estimated as 
$\rm \Delta L/ L_{tot} \simeq 0.04$. For a disk isentropic in 
z-direction, as discussed in NW1-3, the formula $\rm
PV^{\gamma}=const$ ($\gamma=5/3$) gives a 
relation between the temperature and thickness of a disk: 
$\rm Th^{\gamma-1}=const$. Thus, $\rm \vert\delta T/T 
\vert=(2/3)\delta h/h$. The  flux from the radii between 7.5 and 
8.5 will oscillate with the amplitude $\rm \delta(\Delta L)/\Delta 
L=4\delta T/T = (8/3)\delta H_d/H_d$, and we can write 
$$
\rm {{\delta (\Delta L)}\over {L_{tot}}} \simeq
{0.04} \, {8\over3}{{\delta H_d}\over H_d} \simeq 
0.05 \left({{\delta H_d/H_d}\over{1/2}}\right) , 
\eqno(B7)
$$
which means taht $\sim $ few \% is an upper limit for the flux 
oscillations due to $g$-modes in a disk.

\clearpage
\begin{figure}
\caption{ Dimensionless angular velocity, $\theta$, versus
a dimensionless radius r in the disk. A family of solutions given by
expression (10) (see \S 2.1) is presented for the case of a NS. 
The solid line shows the Keplerian solution described by formula (9). 
The dashed and dash-dotted lines correspond to the adjustment radius 
r$=1.3$ and 1.39, respectively. In our calculation the 
$\gamma$-parameter (see equation [8]) is equal to 15.
\label{Fig.1}}
\end{figure}

\begin{figure}
\caption{ Schematic diagram illustrating the centrifugal barrier (CB)
model. The CB oscillates in vertical, radial and azimuthal directions 
around the vicinity of the adjustment radius (see the text for
details). Some fraction of the soft photons from a disk are radiated
in the observer's direction, while most of the photons from a disk 
illuminate the CB region and get upscattered there to higher energies.
\label{Fig.2}}
\end{figure}

\begin{figure}
\caption{Plot of $\rm f \times m$ (where f is the QPO frequency in Hz,
and  $\rm m = M/M_{\odot}$ is the dimensionless mass of a compact
object) versus $\gamma$-parameter ($\rm \propto {\dot M}$).
\label{Fig.3}}
\end{figure}

\begin{figure}
\caption{Temperature of the CB region as a function of ratio
$\rm H/F$ (where H is the external flux of soft photons from a disk
that illuminate the CB region, and F is the intrinsic energy flux from
the CB region. We performed our calculations for the Thomson optical 
depth of the CB region $\tau_0 = 5$.
\label{Fig.4}}
\end{figure}

\noindent                  
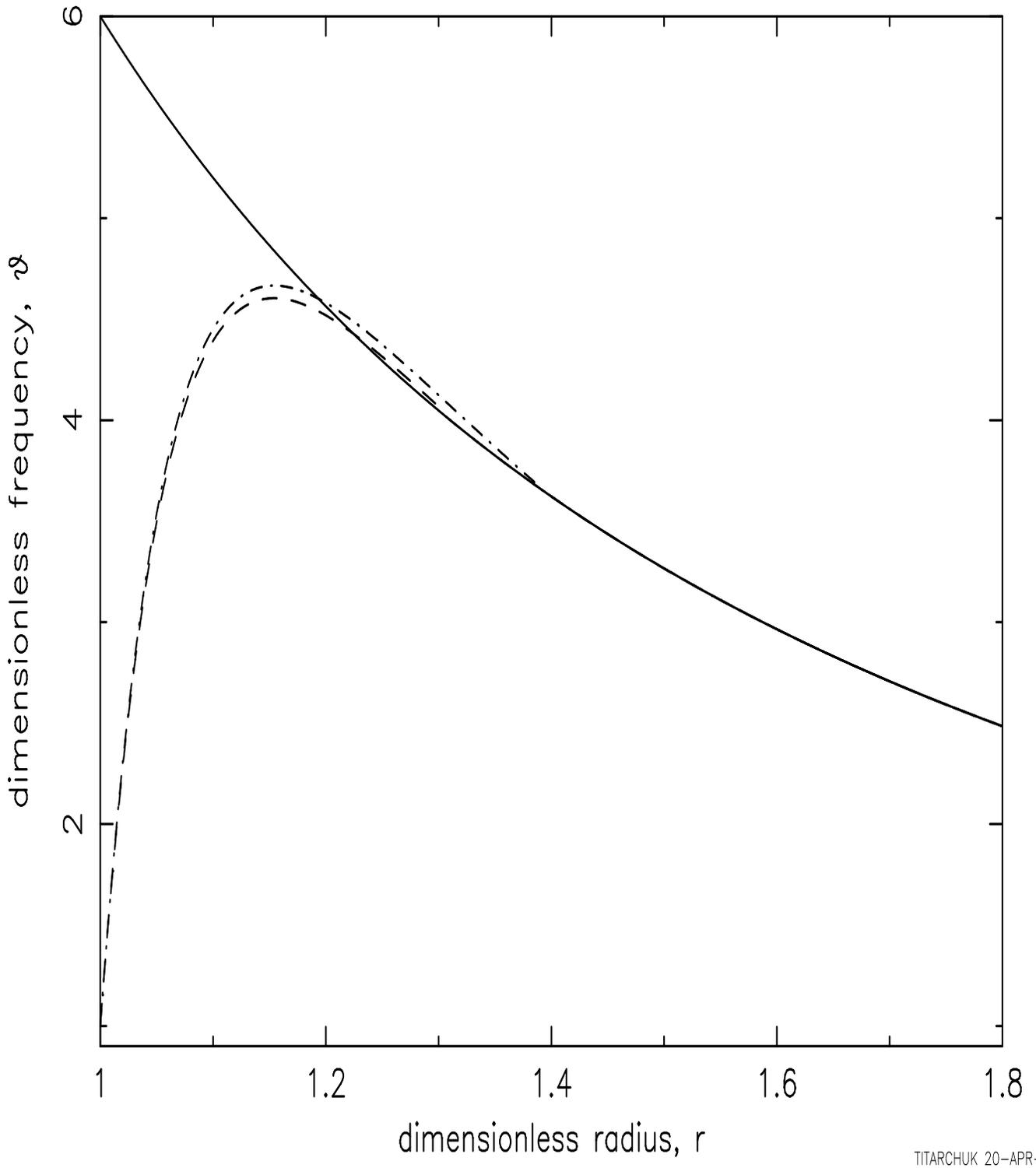
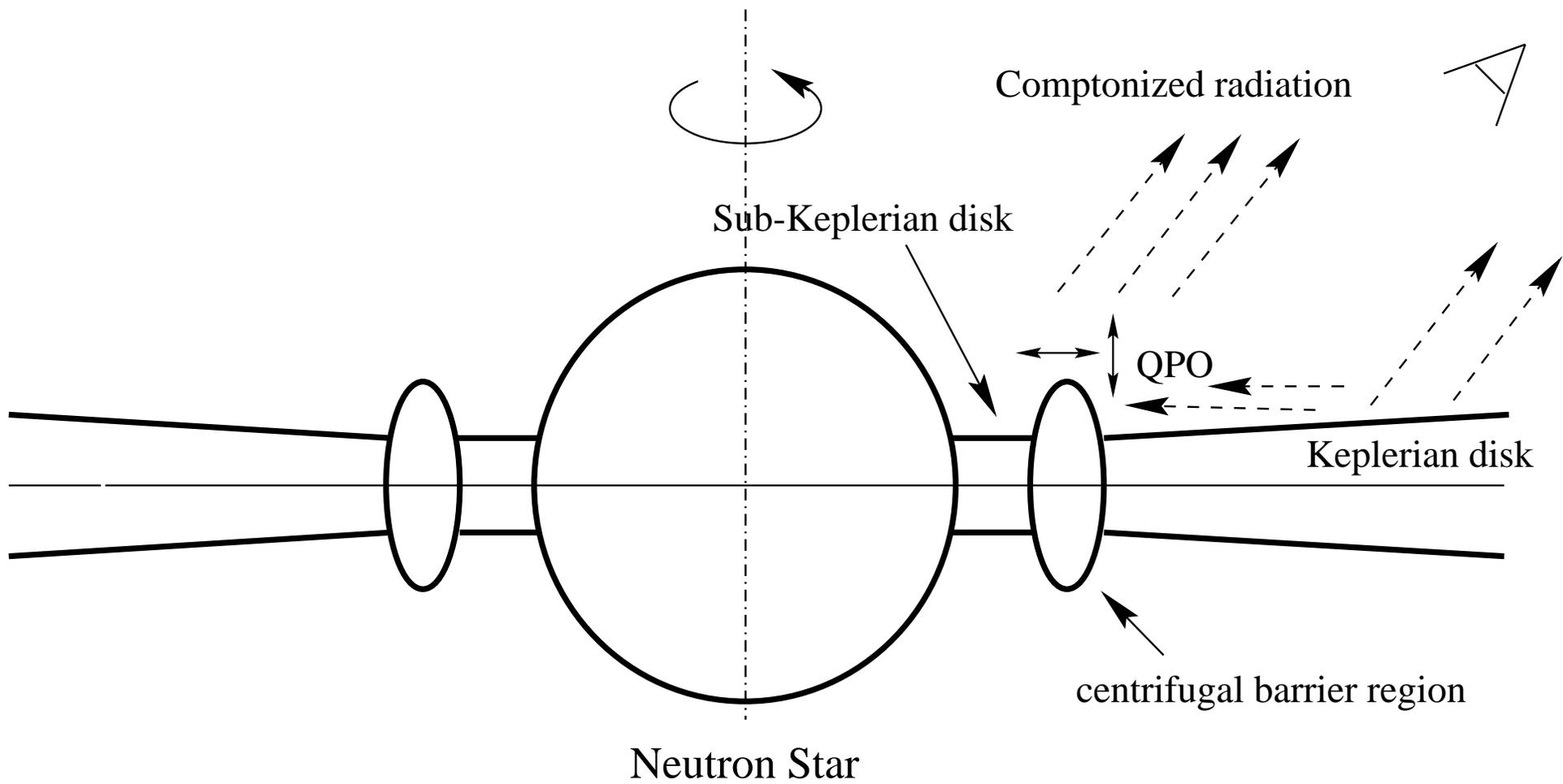
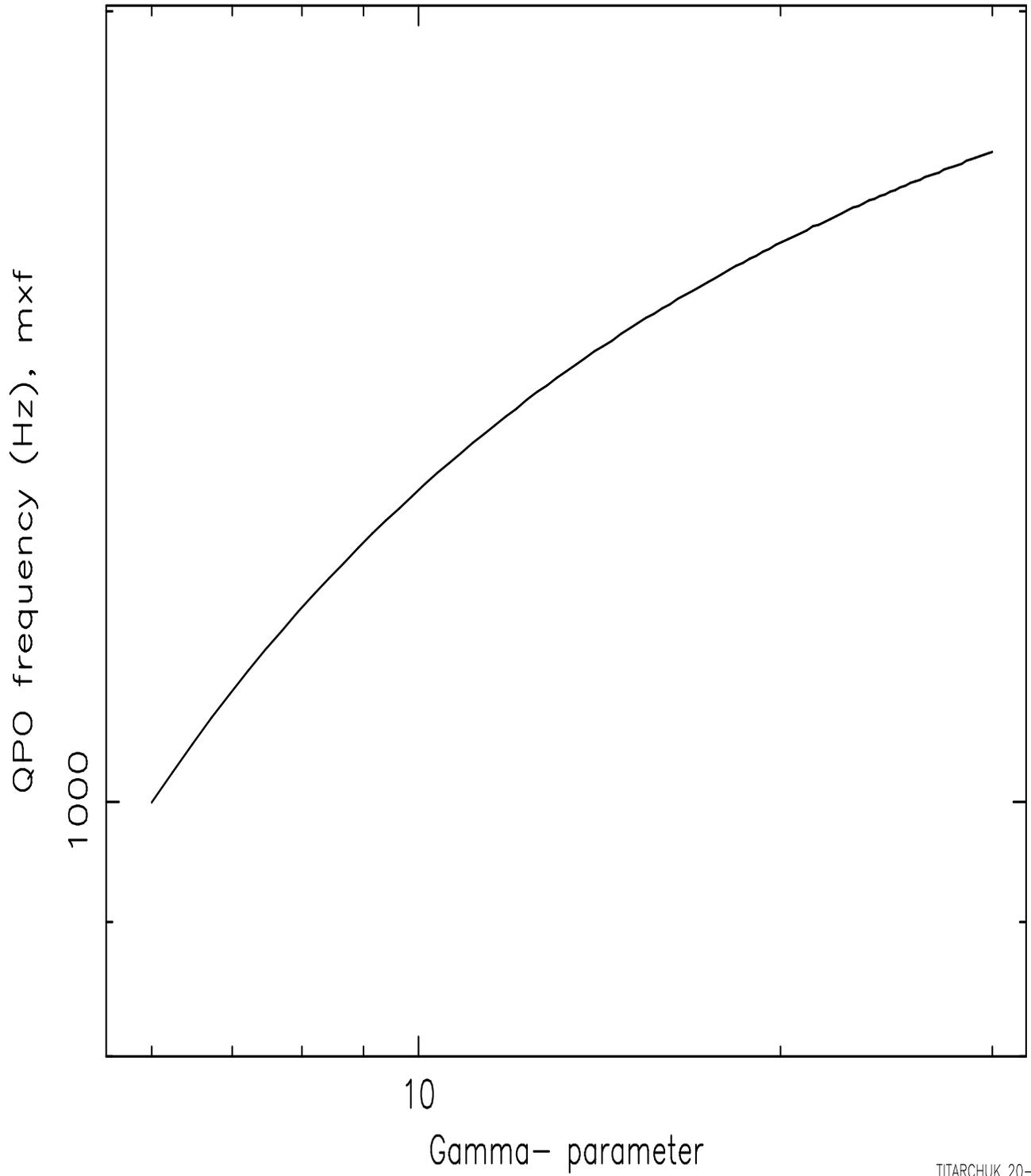
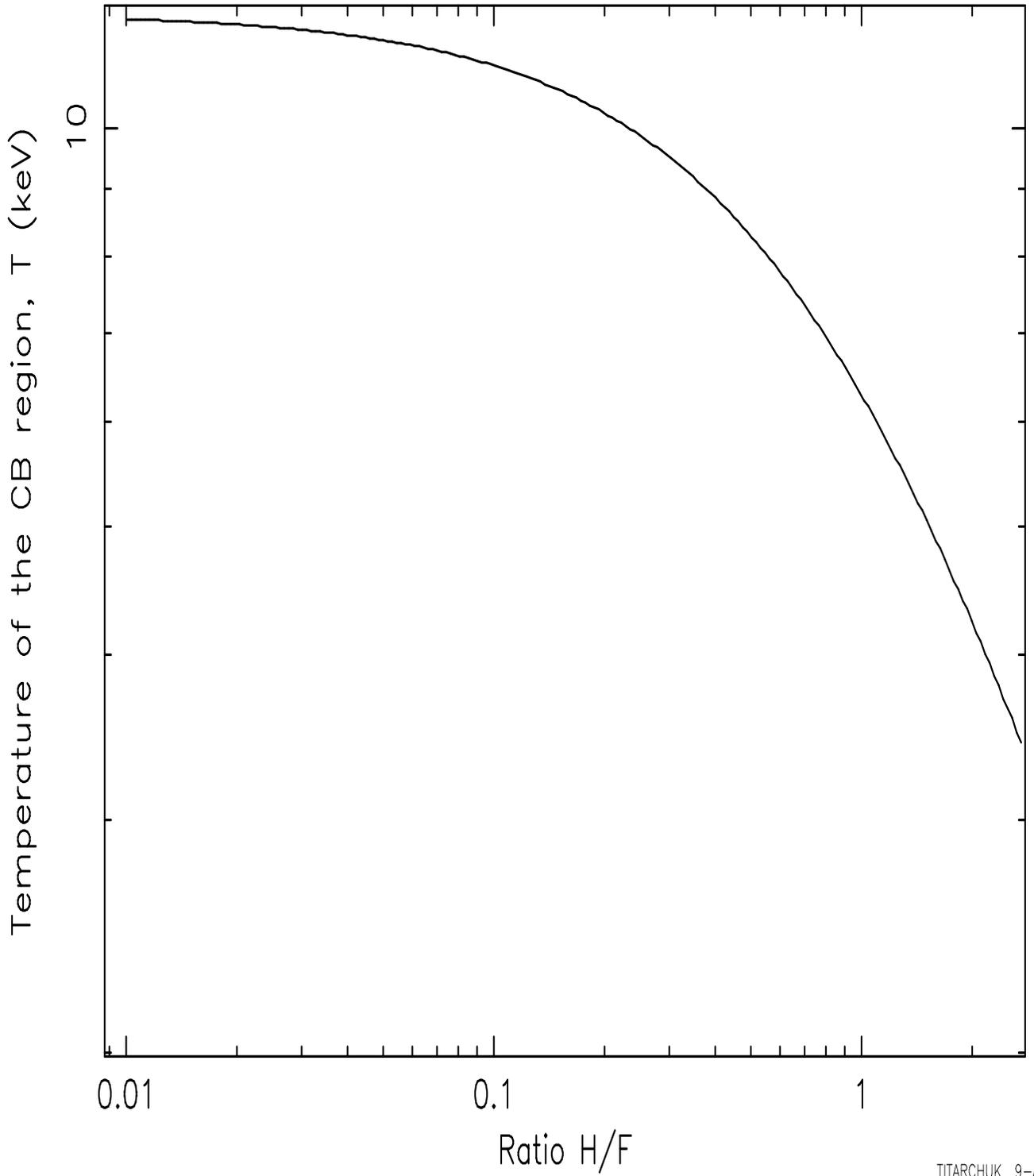
\end{document}